\documentstyle[latexsym]{article}

\def\be{\begin{eqnarray}}
\def\ee{\end{eqnarray}}
\def\nn{\nonumber}

\hoffset= - 1.0in         

\title{{\bf Radiation Beyond Four Space-Time Dimensions
} \vspace{.5cm}}
\author{{\bf A. Mironov}\footnote{E-mail: \ mironov@lpi.ru; mironov@itep.ru}
\date{ } \\
{\small {\it Lebedev Physics Institute}
and {\it ITEP, Moscow, Russia}}\\ \\
{\bf A.Morozov}\thanks{E-mail: \ morozov@itep.ru}
\date{ } \\ {\small {\it ITEP, Moscow, Russia}}
}

\begin{document}

\maketitle

\vspace{-8.0cm}

\begin{center}
\hfill FIAN/TD-14/06\\
\hfill ITEP/TH-77/06\\
\end{center}

\vspace{6.5cm}

\begin{abstract}
\noindent We present a list of formulas describing classical
radiation of the rank $s$ tensor field from an accelerated
point-like source in flat space-time of arbitrary even dimension
$d$. This allows straightforward evaluating the total intensity and
radiated momentum for any $s$ and $d$ algorithmically, by hands or
with the help of a computer (e.g. with an attached MAPLE program).
Practical application of formulas is limited, because, for $s>1$,
the energy-momentum tensor for the point-like source is not
conserved. This usually means that one cannot neglect contributions
to radiation from tensions of the forces that cause acceleration of
the source.
\end{abstract}

\bigskip

\bigskip


Traditionally radiation processes were considered as subjects of
direct physical application and therefore were deeply investigated
only in no more than $d=4$ space-time dimensions, see
\cite{radprocbook}. Even for $d=2$ and $3$, where obvious
applications exist, say, to sound waves in media, the theory remains
badly represented in the literature. Only recently, after the
strings-inspired multi-dimensional models \cite{tsg} attracted
increasing attention \cite{bh}, some papers on multi-dimensional
radiation began to emerge \cite{Kos,multidimrad,turks,mmrad,ccg}.
Of course, they are still too few to cover the field at the level of
exhaustiveness, typical for the literature on $4d$ radiation. In
this paper we make a step, which we think is necessary for a {\it
systematic} study of physical effects. Namely, we provide a list of
generic formulas, describing {\it classical} radiation for arbitrary
dimension $d$ and rank $s$ of the {\it radiated} fields, what should
help to reveal 
the underlying physical and mathematical structures. In particular,
one can immediately read off the radiation damping force in higher
dimensions from our results, for example by the method of
\cite{Kos}.

To illustrate the main results of this paper, we use several {\bf
Tables}. We begin with {\bf Table I}, summarizing the logic of
radiation calculus. The left column is our main line, the right
column enumerates related subjects which deserve deeper physical
consideration and are left for a detailed presentation of radiation
physics in a more sophisticated review article.

According to {\bf Table I}, in order to find radiation intensity one
needs to perform the following chain of calculations:

\bigskip

$\bullet$
Solve wave equation 
for a point-like source of the rank-$s$ field, moving along a world
line $z^\mu(\tau)$, \be \Box A_{\mu_1\ldots\mu_s}(x) = \oint
u_{\mu_1}\ldots u_{\mu_s} \delta^{(d)}\big(x-z(\tau)\big)d\tau, \ee
and pick up the slowest-falling contribution at large distances. For
even $d$ it is given by a simple formula for the retarded
Li\'enard-Wiechert potential: \be A^{rad}_{\mu_1\ldots\mu_s} =
\left(\frac{1}{(Ru)}\,\partial_\tau\right)^\frac{d-4}{2}
\frac{u_{\mu_1}\ldots u_{\mu_s}}{Ru}
\ \ \ {\rm and} \ \ \
\partial_\mu A^{rad}_{\mu_1\ldots\mu_s} =
R_\mu\left(\frac{1}{(Ru)}\,\partial_\tau\right)^\frac{d-2}{2}
\frac{u_{\mu_1}\ldots u_{\mu_s}}{Ru} \label{radpot} \ee where $u_\mu
= \partial_\tau z_\mu$ is source's $d$-velocity, $u^2=1$, evaluated
at the radiation-time moment $t'=z^0(\tau)$ which is determined by
the condition $R^2=0$, where $R_\mu$ is the $d$-vector with
components $R^\mu\equiv x^\mu-z^\mu(\tau)$. Here $\tau$ is source's
proper time. In particular, \be {\partial \tau\over\partial
x^\mu}={R_\mu\over (Ru)}\ee We also introduce the notation
$n^\mu\equiv (1,-\vec n)={R^\mu\over R}$, where $R$ is length of the
spatial part of $R^\mu$, $R\equiv\sqrt{(R^i)^2}$ ($n^\mu$ is {\bf
not} a $d$-vector in contrast to $R^\mu$; we define it in the
laboratory frame).

It is convenient to re-write (\ref{radpot}) in condensed notation:
\be A_{rad} = \frac{1}{R^{d-2}}
\left(\frac{1}{U}\,\partial_\tau\right)^{p-1}\frac{S}{U} =
\frac{1}{R^{d-2}} \sum_{q=0}^{p-1} \big(\partial_\tau^q
S\big)A^{p-1}_q \ \ {\rm and} \ \
\partial_\mu A_{rad} = \frac{n_\mu}{R^{d-2}}
\sum_{q=0}^{p} \big(\partial_\tau^q S\big)A^{p}_q,
\ee
where $U = (nu)$, $S = u_{\mu_1}\ldots u_{\mu_s}$, $p=\frac{d-2}{2}$,
and find $A^p_q$ from recurrent relations \be A^{p+1}_q =
\frac{1}{U}\Big(\partial_\tau A^p_q + A^p_{q-1}\Big) \ee The first
few $A^p_q$ are listed in {\bf Table II}.

\bigskip

$\bullet$ Check transversality: in the leading order in $1/R$, when
derivatives of $R$ are neglected, \be
\partial^{\mu_s} A^{rad}_{\mu_1\mu_2\ldots\mu_s} =
\left(\frac{1}{(Ru)}\,\partial_\tau\right)^\frac{d-2}{2}
u_{\mu_1}\ldots u_{\mu_{s-1}} \label{trans} \ee does not vanish for
$s\geq 2$, and this is {\it not} cured by subtraction of traces. The
physical reason for non-transversality is the disregard of radiation
from tensions of the forces that cause acceleration of the source:
only taking into account all the tensions makes radiation problem
well-defined for $s\geq 2$. For $s\geq 2$ the formulas of the
present paper provide only a part of the full answer.


\bigskip

$\bullet$ Build up the energy-momentum tensor in the wave zone. It
is equal to \be\label{T} T_{\mu\nu}^{(s)} = R_\mu R_\nu \left\{
\left(\frac{1}{(Ru)}\,\partial_\tau\right)^\frac{d-2}{2}
\frac{u_{\mu_1}\ldots u_{\mu_s}} {Ru}\right\}^2 = \frac{R_\mu
R_\nu}{R^d} \left\{
\sum_{q=0}^p A^p_q\ \partial_\tau^q S
\right\}^2 \ee In applications, certain linear combinations of such
stress tensors arise. For instance, for the scalar waves \be
\label{9}T_{\mu\nu}^{(scalar)} =- T_{\mu\nu}^{(0)} \ee This is
because only the spatial components of all non-zero spin fields have
any physical meaning (e.g. survive in physical gauges), thus giving
rise to the overall minus sign of the kinetic part of the
energy-momentum tensor as compared with the scalar case. For
gravitational waves \be T_{\mu\nu}^{(grav)} = T_{\mu\nu}^{(2)} -
\frac{1}{2}T_{\mu\nu}^{(0)} \label{8}\ee
Similar redefinitions are also needed for higher spins $s>2$. These
linear transformations can be easily made using the results for
$T_{\mu\nu}^{(s)}$ read off from {\bf Table V}.

\bigskip

$\bullet$ Evaluate the radiated momentum flux through the sphere of
radius $R$. It is equal to \be {d{\cal P}_{d|s}^\mu}=-\oint
\left\{T^{\mu i}n_i dx^0 \right\}dS \ee where $dS$ is an
infinitesimal element of this sphere (i.e. integration runs over the
sphere). Then, one finally obtains (using $dx^0=Ud\tau=(nu)d\tau$)
\be\label{P} {d{\cal P}_{d|s}^\mu\over d\tau}=-\oint T^{\mu i}n_i
R^{d-2}U d\Omega_{d-2}=\nn\\= \sum_{q'\!,\,q''=0}^p
\Big(\partial_\tau^{q'}
\big(u_{\mu_1}\ldots u_{\mu_s}\big)
\partial_\tau^{q''}
\big(u^{\mu_1}\ldots u^{\mu_s}\big)\Big) \int n^\mu U A^p_{q'}
A^p_{q''} d\Omega_{d-2}(n) = \sum_{k=0}^{\frac{d-4}{2}}
\big(\partial_\tau^k u^\mu\big) P_k^{(d|s)}(\kappa) \ee where
$d\Omega_{d-2}$ is the solid angle in the $d-1$ space and
$P_k^{(d|s)}$ are certain functions of various scalar products
$\big(\partial_\tau^l u_\nu \partial_\tau^m u^\nu\big)$ with $l+m
\leq d-4-k$.
In particular, the radiated energy loss is ($dt'=u^0d\tau=\gamma
d\tau$)\be \label{P0}\frac{d{\cal P}^0_{d|s}}{dt'} ={1\over\gamma}
\sum_{k=0}^{\frac{d-4}{2}} \big(\partial_\tau^k \gamma\big)
P_k^{(d|s)}(\kappa)\ \stackrel{{\rm for\ constant}\ \gamma =
(1-v^2)^{-1/2}} {\longrightarrow}\ P_0^{(d|s)}(\kappa), \ee
We warn once again that due to effects like (\ref{trans}) and
(\ref{8})-(\ref{9}), this quantity represents only a part of the
full answer for $s\ge 2$. Moreover, as follows from {\bf Table V},
it can be even negative (!): for example, for fields with
asymptotically large rank $s$ the leading contribution to the
radiated momentum, $\frac{d{\cal P}^\mu_{d|s}}{d\tau}\ $ behaves as
$\ -(s\dot u^2)^{{d\over 2}-1}\sim (-)^{d\over 2}\ $, provided $\dot
u^2\le 0$. For $s=0$, this is cured by changing the common sign, see
(\ref{9}). However, for $s\ge 3$ the procedure has to be more
sophisticated.

Sometime also contractions $I_{d|s} = u^\mu \frac{d{\cal
P}^\mu_{d|s}}{d\tau}$ are considered \cite{turks}, but they do not
have any direct physical meaning unless $\frac{d{\cal
P}^\mu_{d|s}}{d\tau} \sim u_\mu$, as happens for $d=4$. In this
latter case, $I_{d|s}=\frac{d{\cal P}^0_{d|s}}{dt'}$.

\bigskip

$\bullet$ Calculate angular integrals over isotropic (light-like)
unit vectors $n^\mu$ in (\ref{P}) (although $n^\mu$ is not a
$d$-vector, these integrals are Lorentz-covariant, since (\ref{P})
is Lorentz-covariant)), \be \int \frac{n^{\mu_1}\ldots
n^{\mu_m}}{(nu)^{d+m-2}}d\Omega_{d-2}(n) \sim {\rm Pr}_d\left(
u^{\mu_1}\ldots u^{\mu_m}\right) \label{angin} \ee They are
expressed through the trace-eliminating projector ${\rm Pr}_d$
(since $n^2=0$ and contraction of any pair of $\mu$-indices at the
l.h.s. should give zero):
$$
\begin{array}{ccc}
{\rm spin}\ s=2: && {\rm Pr}_{d}(u_\mu u_\nu)
= u_\mu u_\nu - \frac{1}{d}\eta_{\mu\nu} \\
{\rm spin}\ s=3: && {\rm Pr}_{d}(u_{\mu_1} u_{\mu_2} u_{\mu_3}) =
u_{\mu_1} u_{\mu_2} u_{\mu_3} -
\frac{1}{d+2}\Big(\eta_{\mu_1\mu_2}u_{\mu_3} +
\eta_{\mu_1\mu_3}u_{\mu_2} + \eta_{\mu_2\mu_3}u_{\mu_1} \Big) \\
{\rm spin}\ s=4: && {\rm Pr}_{d}(u_{\mu_1} u_{\mu_2} u_{\mu_3}
u_{\mu_4}) = u_{\mu_1} u_{\mu_2} u_{\mu_3} u_{\mu_4} - \\ && -
\frac{1}{d+4}\Big(\eta_{\mu_1\mu_2}u_{\mu_3}u_{\mu_4} + \ 5\ {\rm
permutations}\Big)  + \frac{1}{(d+4)(d+2)}
\Big(\eta_{\mu_1\mu_2}\eta_{\mu_3\mu_4} +
\eta_{\mu_1\mu_3}\eta_{\mu_2\mu_4} +
\eta_{\mu_1\mu_4}\eta_{\mu_2\mu_3}\Big)\\
\ldots && \\
\end{array}
$$
For generic $m$ we have:
$$
{\rm Pr}_{d}\big(u_{\mu_1}\ldots u_{\mu_m}\big) = \sum_{k=0}^{[m/2]}
(-)^k \frac{(d+2m-4-2k)!!}{(d + 2m-4)!!}
\left(\overbrace{u_{\mu_1}\ldots u_{\mu_{m-2k}}
\Big(\underbrace{\eta_{\mu_{m-2k+1}\mu_{m-2k+2}}\ldots
\eta_{\mu_{m-1}\mu_m}  + \ldots}_{(2k-1)!!\ {\rm permutations}}
\Big) + \ldots}^{C^{m}_{2k}\ {\rm permutations}}\right)
$$
This projector can be conveniently re-written through generating
function: after contraction with $m$ copies of a $d$-vector $x$
we get
$$
{\rm Pr}_{d}\Big((xu)^m\Big) = \sum_{k=0}^{[m/2]} (-)^k
(2k-1)!!C^m_{2k} \frac{(d+2m-4-2k)!!}{(d + 2m-4)!!}
(xu)^{m-2k}(x^2)^k =
$$ \vspace{-0.25cm}
\be
= \sum_{k=0}^{[m/2]} (-)^k
\frac{(d +2m-4-2k)!!}{(d + 2m-4)!!}
\frac{m! (xu)^{m-2k}(x^2)^k}{2^k k! (m-2k)!}
\label{proje}
\ee
The sum of coefficients in this formula is needed to define the
normalization factor in (\ref{angin}): \be c_d(m) =
\sum_{k=0}^{[m/2]} (-)^k C^{m}_{2k}(2k-1)!! \frac{(d+2m-4-2k)!!}{(d
+ 2m-4)!!} = \frac{\Big(d+
2[m/2]-3\Big)!!\Big(d+2(m-[m/2])-4\Big)!!} {(d-3)!!(d+2m-4)!!} \ee
Here $[m/2]$ means integer part of a number, e.g. $[3/2]=1$ and
$3-[3/2] = 2$.

Rewritten in terms of the generating function and with normalization
factor restored, the angular integral, (\ref{angin}) is (see {\bf
Table III}):
$$
\int \frac{(nx)^md\Omega_{d-2}(n)}{(nu)^{d+m-2}} =
\frac{S_{n-2}}{c_d(s)}{\rm Pr}_d\Big((ux)^m\Big) =
$$ \vspace{-0.35cm}
\be
=\frac{2^{d/2}\pi^{\frac{d-2}{2}}}{(d+2[m/2]-3)!!}
\sum_{j=0}^{[m/2]} (-)^j\frac{(d+2m-4-2j)!!}{(d+2m - 2[m/2]-4)!!}
\frac{m!}{2^j j!(m-2j)!} (ux)^{m-2j}(x^2)^j
\label{intgen}
\ee
It remains to substitute the vector $x$ by $x = tx_0 +
\sum_{k=0}^{\frac{d-2}{2}} a_k \partial_\tau^k u$ and pick up the
coefficient in front of the relevant combination of $t$ and $a_k$.
Note that this calculation uses the generation function which
provides an additional combinatorial coefficient that has to be
taken into account manifestly. E.g., in order to get the correct
coefficients in front of the term
$\partial_\tau^{k_1}u\ldots\partial_\tau^{k_n}u$ with different
values of $k_1,\ldots, k_n$, one has to divide the generating
function by $n!$.

The final results can be equally expressed via scalar products
$U_{ij}\equiv\partial_\tau^l \vec u_\nu
\partial_\tau^m \vec u^\nu$
and through the Frenet curvatures $\kappa_m$ and their derivatives
$\partial_\tau^l\kappa_m$. Hence, the last step:

\bigskip

$\bullet$ Re-express $U_{ij}$ through Frenet curvatures
\cite{difgem} which parameterize the moving orthonormal basis,
associated with the world line $x^\mu(\tau)$. It is formed by
$d$-vectors $\vec N^{(\mu)}$, $\mu = 0,\ldots,d-1$, \be \vec
N^{(\mu)}\vec N^{\nu)} = \eta^{\mu\nu} \label{orthon} \ee with $\vec
N^{(0)} = \vec u$ (i.e. $N_\mu^{(0)} = u_\mu$) and other vectors
constructed recursively from the condition
$$
\partial_\tau \vec N^{(\mu)} =
- k_{\mu+1}\vec N^{(\mu+1)} + \sum_{\nu=1}^\mu \beta_{\mu\nu}\vec
N^{(\nu)}
$$
From $\tau$-derivative of orthonormality condition,
$$
\Big(\partial_\tau \vec N^{(\mu)}\Big) \vec N^{(\nu)} + \vec
N^{(\mu)} \Big(\partial_\tau \vec N^{(\nu)}\Big) = 0\ .
$$
It follows that at $\nu\leq\mu$
$$
\beta_{\mu\nu} = \Big(\partial_\tau \vec N^{(\mu)}\Big) \vec
N^{(\nu)} = -\vec N^{(\mu)} \Big(\partial_\tau \vec N^{(\nu)}\Big) =
\kappa_\mu\delta_{\nu,\mu-1}\eta^{\mu\mu}
$$
and \be
\partial_\tau \vec N^{(\mu)} =
- k_{\mu+1}\vec N^{(\mu+1)} + k_\mu\vec N^{\mu-1} \ee Parameters
$k_\mu$ (they are not $d$-vectors!) depend on the shape of the world
line $x^\mu(\tau)$ in the infinitesimal vicinity of its point and
are called Frenet curvatures. Expressions of scalar products
$U_{ij}$ through these curvatures are given in {\bf Table IV}.

\bigskip

%
%
%
%

\section*{Results}

\bigskip

All these steps are easily made with the help of MAPLE or
Mathematica, see Appendix at the end of this paper for a MAPLE
program, \cite{maple}. As an application of the algorithm we obtain
the explicit formulas for the radiated $d$-momentum $d{\cal
P}_{d|s}^\mu$ as well as somewhat simpler formulas for
$P_0^{(d|s)}$, the radiated intensity at constant $\gamma$, (i.e.
$\kappa_2^2={\gamma^2\kappa_1^2\over\gamma^2-1}=$const,
$\kappa_i=0$, $i>2$) see (\ref{P0}), and for the contractions
$I_{d|s} = u^\mu \frac{d{\cal P}^\mu_{d|s}}{d\tau}$, \cite{turks}
for ``realistic" values of $d=4,6,8,10$ and generic rank $s$. First,
we list results for the case of $d=4,6,8$, when the formulas are
relatively simple, and one can understand their general structure.
Much more involved formulas in the most interesting case of $d=10$
are put into a separate section.

\bigskip

\newpage

\framebox{$d{\cal P}_{d|s}^\mu$:}


$$ {d{\cal P}^\mu_{4|s}\over d\tau}= {4-12s\over 3}\pi\dot u^2
u^\mu$$

\bigskip

$$ {d{\cal P}_{6|s}^\mu\over d\tau} =\pi^2\left[{19\over
3}-(2s-3)^2\right]\dot u^4 u^\mu+{8\pi^2\over 15}(1-5s)\ddot u^2
u^\mu+{16\pi^2\over 35}(2-7s)(\dot u\ddot u)\dot u^\mu+{16\pi^2\over
105}(7s-4)\dot u^2\ddot u^\mu$$

\vspace{0.7cm}

$${d{\cal P}_{8|s}^\mu\over d\tau} =\left\{-{16\pi^3\over
9}\left({7s^3\over 10}-{135\over 14}s^2+{838\over 35}s-5\right)\dot
u^6+{32\pi^3\over 3}\left(-s^2+{101\over 35}s-{41\over
63}\right)(\dot u\ddot u)^2+\right.$$

$$
+\left.{16\pi^3\over 3}\left(-{s^2\over 5}+{59\over 35}s-{20\over
63}\right)\dot u^2\ddot u^2+32\pi^3\left({s^2\over 15}-{2s\over
7}+{11\over 189}\right)\dot u^2(\dot u\dot{\ddot u})-{16\pi^3\over
15}\left(s-{1\over 7}\right)\dot{\ddot
u}^2\right\}u^\mu+$$

$$+\left\{{32\pi^3\over 7}\left(-{6s^2\over 5}+{19\over 5}s-{283\over
297}\right)\dot u^2(\dot u\ddot u)+{64\pi^3\over 7}\left({1\over
27}-{s\over 5}\right)\ddot u\dot{\ddot u}\right\}\dot u^\mu
+$$

$$+\left\{{64\pi^3\over 21}\left({2s^2\over 5}-2s+{65\over
99}\right)\dot u^4+{64\pi^3\over 21}\left({1\over 9}-{2s\over
5}\right)\dot u\dot{\ddot u}\right\}\ddot u^\mu+{32\pi^3\over
35}\left(s-{17\over 27}\right)(\dot u\ddot u)\dot{\ddot u}^\mu $$

\vspace{1cm}

\framebox{$P_0^{(d|s)}$}

$$ P_0^{(4|s)}=\frac{12s-4}{3}\pi\kappa_1^2 $$

\bigskip

$$ P_0^{(6|s)}= -\pi^2\left(4s^2-{28\over 3}s+{32\over
15}\right)4\kappa_1^4+ {8\pi^2\over 15}(5s-1)\kappa_1^2\kappa_2^2
$$

\bigskip

$$ P_0^{(8|s)}={16\pi^3\over
105}(7s-1)(\kappa_1^2\kappa_2^2\kappa_3^2+\kappa_1^2\kappa_2^4)+
{8\pi^3\over
315}(49s^3-549s^2+1004s-216)\kappa_1^6-16\pi^3\left({s^2\over
5}-s+{64\over 315}\right)\kappa_1^4\kappa_2^2 $$

\vspace{1cm}

\framebox{$I_{d|s}$}

$$ I_{4|s} = \frac{12s-4}{3}\pi\kappa_1^2 $$

\bigskip

$$ I_{6|s}=-\pi^2\left(4s^2-{124\over 15}s+{32\over
21}\right)\kappa_1^4 + {8\pi^2\over 15}(5s-1)(\dot\kappa_1^2 +
\kappa_1^2\kappa_2^2)$$

\bigskip

$$ I_{8|s}={16\pi^3\over 105}\left[\left({49\over 6}s^3-{167\over
2}s^2+{406\over 3}s-{276\over 11}\right)\kappa_1^6-
\left(21s^2-97s+{172\over 9}\right)\kappa_1^4\kappa_2^2-
\left(77s^2-180s+{109\over 3}\right)\kappa_1^2\dot\kappa_1^2+
\right. $$ $$ \left.\phantom{1\over 8}+
2(7s^2-9s+4)\kappa_1^3\ddot\kappa_1+(7s-1)\left\{\kappa_1^2
\kappa_2^2\kappa_3^2+(\kappa_1\kappa_2^2- \ddot\kappa_1)^2+
(2\kappa_2\dot\kappa_1+ \kappa_1\dot\kappa_2)^2\right\}\right] $$


\section*{Results for the most interesting case of $d=10$}

\bigskip

$${d{\cal P}_{10|s}^\mu\over d\tau} =\left\{{4\pi^4\over
315}\left(19s^4-{6466\over 9}s^3+{236237\over 33}s^2-{1609666\over
99}s+{9800\over 3}\right)\dot u^8-{32\pi^4\over
7}\left(s^3-{200\over 9}s^2+{2147\over 33}s-{6230\over
429}\right)\dot u^2 (\dot u\ddot u)^2+\right.$$

$$\left.+{16\pi^4\over 35}\left(s^3-{61\over 27}s^2 - {8316\over
297}s+{11690\over 1287}\right)\dot u^4\ddot u^2+{64\pi^4\over
35}\left(s^3-{1037\over 54}s^2+{32653\over 594}s-{4865\over
429}\right)\dot u^4(\dot u\dot{\ddot u})-\right.$$

$$\left.-{16\pi^4\over 105}\left(19s^2-{611\over 9}s+{512\over
33}\right)\ddot u^4-{32\pi^4\over 15}\left({17\over
7}s^2-151s+{34\over 21}\right)(\dot u\dot{\ddot u})^2-{64\pi^4\over
105}\left(s^2-{53\over 3}s+{100\over 33}\right)(\dot u\ddot u)(\ddot
u\dot{\ddot u})-\right.$$

$$\left.-{256\pi^4\over 35}\left(s^2-{263\over 108}s+{119\over
198}\right)\ddot u^2(\dot u\dot{\ddot u})-{32\pi^4\over
105}\left(s^2-{134\over 9}s+{25\over 11}\right)\dot u^2\dot{\ddot
u}^2+{64\pi^4\over 105}\left(s^2-{74\over 9}s+{15\over
11}\right)\dot u^2(\ddot u\ddot{\ddot u})+\right.$$

$$\left.+{64\pi^4\over 315}(9s^2-49s+10)(\dot u\ddot u)(\dot
u\ddot{\ddot u})+{32\pi^4\over 105}\left({1\over
9}-s\right)\ddot{\ddot u}^4\right\}u^\mu-\left\{{32\pi^4\over
63}\left({8s^3\over 3}-{593\over 11}s^2+{72173\over 429}s-{5670\over
143}\right)\dot u^4(\dot u\ddot u)+\right.$$

$$\left.+{64\pi^4\over 189}\left(20s^2-{829\over 11}s+{2664\over
143}\right)\ddot u^2(\dot u\ddot u)+{256\pi^4\over
189}\left(5s^2-{133\over 11}s+{453\over 143}\right)(\dot u\dot{\ddot
u})(\dot u\ddot u)-{128\pi^4\over 189}\left(s^2-{109\over
22}s+{147\over 143}\right)\dot u^2(\dot u\ddot{\ddot u})+\right.$$

$$\left.+{320\pi^4\over 819}\left(1-{208\over 33}s\right)\dot
u^2(\ddot u\dot{\ddot u})+{64\pi^4\over 189}\left(2s-{3\over
11}\right)(\dot{\ddot u}\ddot{\ddot u})\right\}\dot
u^\mu+\left\{{128\pi^4\over 189}\left(s^3-{785\over
44}s^2+{30843\over 572}s-{2065\over 143}\right)\dot u^6+\right.$$

$$\left.+{320\pi^4\over 189}\left(s^2-{53s\over 11}+{216\over
143}\right)(\dot u\ddot u)^2-{640\pi^4\over
189}\left(s^2-4s+{1569\over 1430}\right)\dot u^2(\dot u\dot{\ddot
u})-{64\pi^4\over 189}\left(4s^2+s+{123\over 143}\right)\dot
u^2\ddot u^2+\right.$$

$$\left.+{128\pi^4\over 189}\left({2\over 11}-s\right)(\ddot
u\ddot{\ddot u})\right\}\ddot u^\mu+\left\{{256\pi^4\over
4158}\left(22s^2-167s+{795\over 13}\right)\dot u^2(\dot u\ddot
u)+{64\pi^4\over 189}\left({3\over 11}-s\right)(\dot u\ddot{\ddot
u})\right\}\dot{\ddot u}^\mu-$$

$$ -\left\{{32\pi^4\over 945}\left(s^2-{141\over 11}s+{1020\over
143}\right)\dot u^4-{64\pi^4\over 315}\left(s-{23\over
33}\right)\ddot u^2 -{256\pi^4\over 945}\left(s-{29\over
44}\right)(\dot u\dot{\ddot u})\right\}\ddot{\ddot u}^\mu $$

\vspace{.5cm}

\rule{7cm}{.5pt}

\vspace{.5cm}

$$
P_0^{(10|s)}={32\pi^4\over 315}\left[\left(-{19\over
8}s^4+{2747\over 36}s^3-{152021\over 264}s^2+{369827\over
396}-192\right)\kappa_1^8+\left({9\over
2}s^3-605s^2+933s-188\right)\kappa_1^6\kappa_2^2- \right.
$$

$$\left.-\left({33\over
2}s^2-325s+28\right)\kappa_1^4\kappa_2^4-\left(9s^2-{88\over
3}s+{43\over 3}\right)\kappa_1^4\kappa_2^2\kappa_3^2+
\left(3s-{1\over
3}\right)(\kappa_1^2\kappa_2^6+\kappa_1^2\kappa_2^2\kappa_3^2\kappa_4^2+
2\kappa_1^2\kappa_2^4\kappa_3^2+\kappa_1^2\kappa_2^2\kappa_3^4)\right]
$$

\vspace{.5cm}

\rule{7cm}{.5pt}

\vspace{.5cm}

$$
I_{10|s}={16\pi^4\over 945}\left[\left({81}s^3-{1693}s^2+{53244\over
11}s-{133048\over
143}\right)\kappa_1^6\kappa_2^2+\left(243s^3-{4785}s^2+{106818\over
11}s-{300280\over 143}\right)\kappa_1^4\dot\kappa_1^2-\right.
$$

$$
-\left({99}s^2-{931}s+{1744\over 11}\right)
\kappa_1^4\kappa_2^4+\left(36s^2-{220}s+{416\over
11}\right)\kappa_1^4\kappa_2\ddot\kappa_2+\left(324s^2-1780s+{3392\over
11}\right)\kappa_1^3\ddot\kappa_1\kappa_2^2- $$

$$
-\left( 468s^2-1160s+{2812\over
11}\right)\kappa_1\ddot\kappa_1\dot\kappa_1^2
+\left(144s^2-568s+{1336\over
11}\right)\kappa_1^2\dot\kappa_1\dot{\ddot\kappa}_1-\left(171s^2-413s+{1062\over
11}\right)\dot\kappa_1^4- $$

$$
-\left(54s^2-488s+{866\over
11}\right)\kappa_1^4\kappa_2^2\kappa_3^2- \left({57\over
4}s^4-{2507\over 6}s^3+{125989\over 44}s^2-{3644975\over
858}s+{111744\over 143}\right)\kappa_1^8-
$$

$$
-\left(108s^3-1506s^2+{31794\over 11}s-{83808\over
143}\right)\kappa_1^5\ddot\kappa_1- \left(324s^2-822s+{1878\over
11}\right) \kappa_1^2\ddot\kappa_1^2-\left(432s^2-2752s+{5760\over
11}\right)\kappa_1^3\dot\kappa_1\kappa_2\dot\kappa_2-
$$

$$
-\left(18s^2-268s+{450\over 11}\right)\kappa_1^4\dot\kappa_2^2
-\left(450s^2-3714s+{7520\over
11}\right)\kappa_1^2\dot\kappa_1^2\kappa_2^2+
$$

$$
+2(9s-1)[(\dot{\ddot\kappa}_1-3\dot\kappa_1\kappa_2^2-
3\kappa_1\kappa_2\dot\kappa_2)^2+(3\ddot\kappa_1\kappa_2+3\dot\kappa_1\dot\kappa_2+
\kappa_1\ddot\kappa_2-\kappa_1\kappa_2^3)^2+(2\kappa_1\dot\kappa_2\kappa_3+
\kappa_1\kappa_2\dot\kappa_3+3\dot\kappa_1\kappa_2\kappa_3)^2-
$$

$$\left.-
\kappa_1^2\kappa_2^2\kappa_3^2(2\kappa_2^2+\kappa_3^2+\kappa_4^2)-
2\kappa_1\kappa_2\kappa_3^2(3\ddot\kappa_1\kappa_2+\kappa_1\ddot\kappa_2+
3\dot\kappa_1\dot\kappa_2)]\right]
$$

\vspace{1cm}

\rule{11cm}{1pt}

\vspace{1cm}

Note that, in the expressions for the radiated momentum the
coefficients in front of terms of degree $k$ in $U_{ij}$ are
polynomials of degree $k$ in rank $s$. Relative simplicity of the
coefficients of these polynomials implies that a general formula may
exist for them at any dimension $d$ and for arbitrary rank $s$.
However, we stress once again that these expressions for higher
ranks are of restricted use, because of necessity to take into
account contributions from other energy-momentum tensors. The second
thing one has to do in order to separate the contribution of the
given {\it spin} $s$ from the expression for the {\it rank} $s$ is
to add proper combinations of energy-momentum tensors for lower rank
fields. This is done straightforward using formulas we obtained
here.

The examples of these formulas, with illustrative purposes, are
collected in {\bf Table V} for low values of $s=0,1,2,3$ for
dimensions $d=4,6,8$ when the formulas are more comprehensible. Much
more involved (but interesting) case of $d=10$ is listed after that
table. The Appendix contains an example of the MAPLE program for
calculating various quantities obtained in this paper.

\section*{Acknowledgements}

We appreciate informative discussions with
P.Kazinsky and S.Lyakhovich. Our work is partly supported by the the
program of the Federal Nuclear Energy Agency, by grants RFBR
07-02-00878-  (A.Mir.) and 07-02-00645 (A.Mor.), joint grant
06-01-92059-CE, by the Grant of Support for the Scientific Schools
NSh-8004.2006.2, by NWO project 047.011.2004.026, INTAS grant
05-1000008-7865 and ANR-05-BLAN-0029-01 project.

\newpage

\part*{Tables}

\vspace{1cm}

\noindent
\begin{tabular}{|ccc|}
\hline &&\\
{\bf TABLE I} && Radiation: calculus and problems \\
&&\\
\hline &&\\
accelerating point-like body (source) && \\
&&\\ $\downarrow$ & $\searrow$ & \\ &&\\
radiation field && separation of radiation from co-moving field, wave zone\\
&&\\ $\downarrow$ & $\searrow$ & \\ &&\\
even dimensions &&
strong violation of Huygens principle at finite distances \\
&& in odd dimensions\\
&&\\ $\downarrow$ & $\searrow$ & \\ &&\\
retarded field \
$A_{rad}^{\mu_1\ldots\mu_s}
= \left(\frac{1}{(Ru)}\partial_\tau\right)^{\frac{d-4}{2}}\!\!
\frac{u^{\mu_1}\ldots u^{\mu_s}}{(Ru)}$ &&
difference between observation and radiation times\\
&&\\ $\downarrow$ & $\searrow$ & \\ &&\\
field tension && transversality violation for $s>1$ \\
$F_\mu^{\mu_1\ldots\mu_s} = A_\mu^{\mu_1\ldots\mu_s} \sim R_\mu
\left(\frac{1}{(Ru)}\partial_\tau\right)^{\frac{d-2}{2}}\!\!
\frac{u^{\mu_1}\ldots u^{\mu_s}}{(Ru)}$
&& $F_{\mu_s}^{\mu_1\ldots\mu_s} \sim
\left(\frac{1}{(Ru)}\partial_\tau\right)^{\frac{d-2}{2}}\!\!
u^{\mu_1}\ldots u^{\mu_{s-1}}$ \\
&& other sources of radiation can not be neglected \\
&&\\ $\downarrow$ & $\searrow$ & \\ &&\\
stress tensor && picking the definite-spin contribution,\\
$T_{\mu\nu} = R_\mu R_\nu \left\{
\left(\frac{1}{(Ru)}\partial_\tau\right)^{\frac{d-2}{2}}\!\!
\frac{u^{\mu_1}\ldots u^{\mu_s}}{(Ru)}\right\}^2$ &&
radiation from other sources to cure $T_{\mu\nu}$ non-consevation\\
&&\\ $\downarrow$ & $\searrow$ & \\ &&\\
full intensity && radiation friction \\
&&\\ $\downarrow$ &&\\ &&\\
polarization, angular and frequency distributions && \\ &&\\
\hline
\end{tabular}

\bigskip

$$
\begin{array}{|l|c|c|c|c|c|}
\hline
&&&&&\\
{\rm {\bf TABLE}\ II:} \ \ A^p_q &q=0&q=1&q=2&q=3&q=4\\
&&&&&\\
\hline
\ \ {\rm field \ in}\ d=4&&&&&\\
p=0\ &\frac{1}{U}& -& -& -&-\\
&&&&&\\
\hline
\ \ {\rm field \ in}\ d=6 &&&&&\\
p=1& -\frac{\dot U}{U^3}& \frac{1}{U^2}& - & -&-\\
\ \ {\rm tension\ in}\ d=4 &&&&&\\
\hline
\ \ {\rm field \ in}\ d=8&&&&&\\
p=2\ &\frac{-U\ddot U + 3\dot U^2}{U^5}&
 - \frac{3\dot U}{U^4} & + \frac{1}{U^3}& - & -\\
\ \ {\rm tension\ in}\ d=6&&&&&\\
\hline
\ \ {\rm field \ in}\ d=10&&&&&\\
p=3\  & \frac{-\dot{\ddot U}U^2 + 10\ddot U\dot U U
- 15\dot U^3}{U^7}&
\frac{-4\ddot U U + 15 \dot U^2}{U^6} & - \frac{6\dot U}{U^5} &
\frac{1}{U^4}& -\\
\ \ {\rm tension\ in}\ d=8&&&&&\\
\hline
\ \ {\rm field \ in}\ d=12&&&&&\\
p=4\  &\frac{-\ddot{\ddot U}U^3 + 15\dot{\ddot U}\dot U U^2
+10\ddot U^2U^2 - 105\ddot U\dot U^2 U + 105\dot U^4}{U^9}
&\frac{-5\dot{\ddot U}U^2 + 60\ddot U\dot U U
- 105\dot U^3}{U^8}&
\frac{-10\ddot U U + 45\dot U^2}{U^7}&-\frac{10\dot U}{U^6}&
\frac{1}{U^5}\\
\ \ {\rm tension\ in}\ d=10&&&&&\\
\hline
\end{array}
$$

\bigskip

$$ \hspace{-2.0cm} {\footnotesize
\begin{array}{|c|c|c|c|c|}
\hline &&&&\\
&{\rm {\bf TABLE}\ III:}&{\rm Formulas}\ (\ref{intgen})
&{\rm for\ angular\ integrals}&\\
&&&&\\
\hline &&&&\\
m& d=4 & d=6 & d=8 & d=10 \\
& (p=0) & (p=1) & (p=2) & (p=3) \\
&&&& \\
\hline
&&&& \\
m=1 &4\pi (ux)&\frac{8\pi^2}{3} (ux)&\frac{16\pi^3}{15} (ux)&\frac{32\pi^4}{105} (ux) \\
&&&& \\
\hline
&&&& \\
m=2 &\frac{4\pi}{3}\Big(4(ux)^2-x^2\Big)&
\frac{8\pi^2}{15}\Big(6(ux)^2-x^2\Big)&
\frac{16\pi^3}{105}\Big(8(ux)^2-x^2\Big)&
\frac{32\pi^4}{945}\Big(10(ux)^2-x^2\Big)\\
&&&& \\
\hline
&&&& \\
m=3 &\frac{4\pi}{3}\Big(6(ux)^3-3(ux)x^2\Big)&
\frac{8\pi^2}{15}\Big(8(ux)^3-3(ux)x^2\Big)&
\frac{16\pi^3}{105}\Big(10(ux)^3-3(ux)x^2\Big)&
\frac{32\pi^4}{945}\Big(12(ux)^3-3(ux)x^2\Big) \\
&&&& \\
\hline
&&&& \\
m=4 &&\frac{8\pi^2}{105}\Big(10\cdot 8(ux)^4-&
\frac{16\pi^3}{945}\Big(12\cdot 10(ux)^4-&
\frac{32\pi^4}{10395}\Big(14\cdot 12(ux)^4-\\
&&-8\cdot 6(ux)^2x^2 + 3(x^2)^2\Big)&
-10\cdot 6(ux)^2x^2 + 3(x^2)^2\Big)&
-12\cdot 6(ux)^2x^2 + 3(x^2)^2\Big)\\
&&&& \\
\hline
&&&& \\
m=5 &&\frac{8\pi^2}{105}\Big(12\cdot 10(ux)^5-&
\frac{16\pi^3}{945}\Big(14\cdot 12(ux)^5-&
\frac{32\pi^4}{10395}\Big(16\cdot 14(ux)^5-\\
&&-10\cdot 10(ux)^3x^2 + 5\cdot 3(ux)(x^2)^2\Big)&
-12\cdot 10(ux)^3x^2 + 5\cdot 3(ux)(x^2)^2\Big)&
-14\cdot 10(ux)^3x^2 + 5\cdot 3(ux)(x^2)^2\Big)\\
&&&& \\
\hline
&&&& \\
m=6 &&&\frac{16\pi^3}{10395}
\Big(16\cdot 14\cdot 12(ux)^6- &
\frac{32\pi^4}{135135}
\Big(18\cdot 16\cdot 14(ux)^6- \\
&&&-14\cdot12\cdot 15(ux)^4x^2 +& -16\cdot14 \cdot 15(ux)^4x^2 +\\
&&&+12\cdot 15\cdot 3(ux)^2(x^2)^2 - 15(x^2)^3\Big)&
+14\cdot 15\cdot 3(ux)^2(x^2)^2 -  15(x^2)^3\Big) \\
&&&& \\
\hline
&&&& \\
m=7 &&&\frac{16\pi^3}{10395}
\Big(18\cdot 16\cdot 14(ux)^7- &
\frac{32\pi^4}{135135}
\Big(20\cdot 18\cdot 16(ux)^7- \\
&&&-16\cdot14\cdot 21(ux)^5x^2 +& -18\cdot16\cdot 21(ux)^5x^2 +\\
&&&+14\cdot 35\cdot 3(ux)^3(x^2)^2 -  15(ux)(x^2)^3\Big)&
+16\cdot 35\cdot 3(ux)^3(x^2)^2 -  15(ux)(x^2)^3\Big)\\
&&&& \\
\hline
&&&& \\
m=8 &&&& \frac{32\pi^4}{2027025}
\Big(22\cdot 20\cdot 18\cdot 16(ux)^8-\\
&&&& - 20\cdot 18\cdot16\cdot 28(ux)^6x^2 + \\
&&&& + 18\cdot 16\cdot 70\cdot 3(ux)^4(x^2)^2 -\\
&&&& - 16\cdot 28\cdot15(ux)^2(x^2)^3 + 105(x^2)^4\Big)\\
&&&& \\
\hline
&&&& \\
m=9&&&& \frac{32\pi^4}{2027025}
\Big(24\cdot 22\cdot 20\cdot 18(ux)^9- \\
&&&& -22\cdot 20\cdot18\cdot 36(ux)^7x^2 +\\
&&&& + 20\cdot 18\cdot 126\cdot 3(ux)^5(x^2)^2 -\\
&&&& -18\cdot 84\cdot15(ux)^3(x^2)^3 + \\
&&&& + 9\cdot 105(ux)(x^2)^4\Big)\\
&&&& \\
\hline
\end{array}
}\hspace{2.0cm}
$$

\bigskip


$$\hspace{-1.3cm}
\begin{array}{|c|c|c|}
\hline &&\\
{\rm {\bf TABLE}\ IV:}& {\rm Scalar\ products\ through\ Frenet\
curvatures}& {\rm Unit\ vectors}\ {\vec N}^{(\nu)},
\ \ {\vec N}^{(\mu)}{\vec N}^{(\nu)} = \eta^{\mu\nu} \\
&&\\ \hline &&\\
{\rm order}\ 0 & u^2 = 1 &  {\vec N}^{(0)} = {\vec u}, \ \ \ \
\dot {\vec N}^{(0)} = \kappa_1{\vec N}^{(1)} \\
&&\\ \hline &&\\
{\rm order}\ 1 & u\dot u = 0 & \\
&  \dot u^2 = -\kappa_1^2& \vec N^{(1)} = \frac{\dot {\vec
u}}{\kappa_1}, \ \ \ \ \dot {\vec N}^{(1)} = \kappa_1 {\vec N}^{(0)}
-
\kappa_2{\vec N}^{(2)} \\
&&\\ \hline &&\\
{\rm order}\ 2 & u\ddot u = \kappa_1^2 &
 \\
& \dot u\ddot u = -\kappa_1\dot\kappa_1 &{\vec N}^{(2)} =
\frac{1}{\kappa_1\kappa_2}\left(-\ddot {\vec u}+
\frac{\dot\kappa_1}{\kappa_1}\dot {\vec u} + \kappa_1^2 {\vec
u}\right)
\\
&\ddot u^2 = \kappa_1^4 - \kappa_1^2\kappa_2^2 -\dot\kappa_1^2 &\dot
{\vec N}^{(2)}=\kappa_2{\vec N}^{(1)}-
\kappa_3{\vec N}^{(3)}\\
&&\\ \hline &&\\
{\rm order}\ 3 &  u\dot{\ddot u} = 3\kappa_1\dot\kappa_1& \\
& \dot u \dot{\ddot u} = -\kappa_1\ddot\kappa_1 - \kappa_1^4 +
\kappa_1^2\kappa_2^2 & \vec N^{(3)} =
\frac{1}{\kappa_1\kappa_2\kappa_3}\Big(\dot{\ddot {\vec u}} -
\big(\frac{2\dot\kappa_1}{\kappa_1}+
\frac{\dot\kappa_2}{\kappa_2}\big)\ddot {\vec u} +\\
&\ddot u \dot{\ddot u} = -\dot\kappa_1 \ddot \kappa_1 +
2\kappa_1^3\dot\kappa_1 - \kappa_1\kappa_2^2\dot\kappa_1 -
\kappa_1^2\kappa_2\dot\kappa_2&+\big(\kappa_2^2-\kappa_1^2 +
\frac{2\dot\kappa_1^2}{\kappa_1^2} +
\frac{\dot\kappa_1\dot\kappa_2}{\kappa_1\kappa_2}
-\frac{\ddot\kappa_1}{\kappa_1} \big) \dot {\vec u}
+\kappa_1^2\big(\frac{\dot\kappa_2}{\kappa_2}-
\frac{\dot\kappa_1}{\kappa_1}\big){\vec u}\Big)\\
&\dot{\ddot u}^2 = -\ddot\kappa_1^2 - 2\kappa_1^3\ddot\kappa_1 +
2\kappa_1\kappa_2^2\ddot\kappa_1 + 9\kappa_1^2\dot\kappa_1^2 -
4\kappa_2^2\dot\kappa_1^2 - &\dot {\vec N}^{(3)}=
\kappa_3{\vec N}^{(2)}-\kappa_4{\vec N}^{(4)} \\
&- 4\kappa_1\kappa_2\dot\kappa_1\dot\kappa_2 -
\kappa_1^2\dot\kappa_2^2 - \kappa_1^6 + 2\kappa_1^4\kappa_2^2 -
\kappa_1^2\kappa_2^4
- \kappa_1^2\kappa_2^2\kappa_3^2&\\
&&\\ \hline &&\\
{\rm order}\ 4 & u\ddot{\ddot u} = 4\kappa_1\ddot\kappa_1 +
3\dot\kappa_1^2 + \kappa_1^4 - \kappa_1^2\kappa_2^2&\\
&\dot u \ddot{\ddot u} = -6\kappa_1^3\dot\kappa_1 +
3\kappa_1\kappa_2^2\dot\kappa_1 + 3\kappa_1^2\kappa_2\dot\kappa_2
- \kappa_1\dot{\ddot \kappa_1}&\\
&\ddot u \ddot{\ddot u} = - \dot\kappa_1\dot{\ddot\kappa}_1
+4\kappa_1^3\ddot\kappa_1 -3\kappa_1\kappa_2^2\ddot\kappa_1 -
\kappa_1^2\kappa_2\ddot\kappa_2 - & \vec N^{(4)}=
{1\over\kappa_1\kappa_2\kappa_3\kappa_4}\left[-\ddot{\ddot{\vec u}}+
\left(3{\dot\kappa_1\over\kappa_1}+2{\dot\kappa_2\over\kappa_2}
{\dot\kappa_3\over\kappa_3}\right)\dot{\ddot{\vec u}}+\right.\\
& -3\kappa_1^2\dot\kappa_1^2 +3\kappa_2^2\dot\kappa_1^2 + \kappa_1^6
- 2\kappa_1^4\kappa_2^2 + \kappa_1^2\kappa_2^4 +
\kappa_1^2\kappa_2^2\kappa_3^2&+\left(\kappa_1^2-\kappa_2^2-\kappa_3^2
+3{\ddot\kappa_1\over\kappa_1}-6{\dot\kappa_1^2\over\kappa_1^2}-
2{\dot\kappa_2^2\over\kappa_2^2}-\right.\\
&\dot{\ddot u}\ddot{\ddot u}=-\ddot
\kappa_1\dot{\ddot\kappa}_1+6\kappa_1^2\dot\kappa_1\ddot\kappa_1-
\kappa_1\kappa_2^2\kappa_3^2\dot\kappa_1-
\kappa_1^2\kappa_2\kappa_3^2\dot\kappa_2-&
\left.-4{\dot\kappa_1\dot\kappa_2\over\kappa_1\kappa_2}
-2{\dot\kappa_1\dot\kappa_3\over\kappa_1\kappa_3}-{\dot\kappa_2\dot\kappa_3
\over\kappa_2\kappa_3}\right)\ddot{\vec
u}+\left({\dot{\ddot\kappa}_1\over\kappa_1}+6{\dot\kappa_1^3\over\kappa_1^3}+
\right.\\&-\kappa_1^2\kappa_2^2\kappa_3\dot\kappa_3
-3\kappa_2^2\dot\kappa_1\ddot\kappa_1+9\kappa_1\dot\kappa_1^3+
4\kappa_1^3\kappa_2^2\dot\kappa_1+2\kappa_1^4\kappa_2\dot\kappa_2-&
+(\kappa_1^2+\kappa_2^2+\kappa_3^2){\dot\kappa_1\over\kappa_1}-
(2\kappa_1^2+\kappa_2^2){\dot\kappa_2\over\kappa_2}-(\kappa_1^2-
\kappa_2^2){\dot\kappa_3\over\kappa_3}-
\\&
-\kappa_1\kappa_2^4\dot\kappa_1-2\kappa_1^2\kappa_2^3\dot\kappa_2+
\kappa_1\kappa_2^2\dot{\ddot\kappa}_1-3\kappa_1^5\dot\kappa_1-
\kappa_1^3\dot{\ddot\kappa}_1-&
-6{\dot\kappa_1\ddot\kappa_1\over\kappa_1^2}-{\dot\kappa_1\ddot\kappa_2\over
\kappa_1\kappa_2}-2{\ddot\kappa_1\dot\kappa_2\over\kappa_1\kappa_2}-
{\ddot\kappa_1\dot\kappa_3\over\kappa_1\kappa_3}+4{\dot\kappa_1^2\dot\kappa_2
\over\kappa_1^2\kappa_2}+2{\dot\kappa_1\dot\kappa_2^2\over\kappa_1\kappa_2^2}+
\\&-2\kappa_2\dot\kappa_1\kappa_1\ddot\kappa_2-
6\kappa_2\dot\kappa_1^2\dot\kappa_2-3\kappa_1\dot\kappa_1\dot\kappa_2^2-
\kappa_1^2\dot\kappa_2\ddot\kappa_2&
\left.+2{\dot\kappa_1^2\kappa_3\over\kappa_1^2\kappa_3}
+{\dot\kappa_1\dot\kappa_2\dot\kappa_3\over\kappa_1\kappa_2\kappa_3}
\right)\dot{\vec u}+\kappa_1^2\left({\ddot\kappa_1\over\kappa_1}-
{\ddot\kappa_2\over\kappa_2}+\kappa_3^2+\right.
\\
&\ddot{\ddot u}^2=-9\kappa_2^2\kappa_3^2\dot\kappa_1^2-
\kappa_1^2\kappa_2^2\kappa_3^2\kappa_4^2+
6\kappa_2^2\dot\kappa_1\dot{\ddot\kappa}_1-9\kappa_1^2\kappa_2^2\dot\kappa_2^2-&
+\left.\left.2{\dot\kappa_2^2\over\kappa_2^2}-2{\dot\kappa_1\dot\kappa_2\over
\kappa_1\kappa_2}-{\dot\kappa_1\dot\kappa_3\over\kappa_1\kappa_3}+
{\dot\kappa_2\dot\kappa_3\over\kappa_2\kappa_3}\right)\vec u\right]\\
&-14\kappa_1^3\kappa_2^2\ddot\kappa_1-
6\kappa_2^2\kappa_3\dot\kappa_1\kappa_1\dot\kappa_3-
4\kappa_1^2\kappa_3\dot\kappa_2\kappa_2\dot\kappa_3-& \dot {\vec
N}^{(4)} = \kappa_4{\vec N}^{(3)} - \kappa_{5}{\vec N}^{(5)}
\\&
-6\dot\kappa_1\dot\kappa_2\kappa_1\kappa_2\kappa_3^2+
30\kappa_1^2\kappa_2^2\dot\kappa_1^2-
\kappa_1^2\kappa_2^2\kappa_3^4-2\kappa_1^2\kappa_2^4\kappa_3^2-&\\&
-\kappa_1^2\kappa_2^2\dot\kappa_3^2+
2\kappa_1^4\kappa_2^2\kappa_3^2-3\kappa_1^6\kappa_2^2-
30\kappa_1^4\dot\kappa_1^2+3\kappa_1^2\kappa_2^4-&\\&-9\kappa_2^4\dot\kappa_1^2
+30\kappa_1^3\dot\kappa_1\kappa_2\dot\kappa_2-
12\kappa_2^3\dot\kappa_1\kappa_1\dot\kappa_2+8\kappa_1^5\ddot\kappa_1+&\\&
+16\kappa_1^2\ddot\kappa_1^2-9\kappa_2^2\ddot\kappa_1^2-\kappa_1^2\ddot\kappa_2^2+
6\kappa_1\kappa_2\dot\kappa_2\dot{\ddot\kappa}_1-
18\dot\kappa_1\dot\kappa_2\kappa_2\ddot\kappa_1-&\\&
-6\dot\kappa_1\dot\kappa_2\kappa_1\ddot\kappa_2+
6\kappa_1\kappa_2^2\kappa_3^2\ddot\kappa_1+
2\kappa_1^2\kappa_2\kappa_3^2\ddot\kappa_2-
6\kappa_2\ddot\kappa_1\kappa_1\ddot\kappa_2+&\\&
+24\kappa_1\ddot\kappa_1\dot\kappa_1^2-12\kappa_1^2\dot\kappa_1\dot{\ddot\kappa}_1-
2\kappa_1^4\kappa_2\ddot\kappa_2+6\kappa_1\kappa_2^4\ddot\kappa_1+&\\&
+2\kappa_1^2\kappa_2^3\ddot\kappa_2-\dot{\ddot\kappa}_1^2&\\ &&\\
\hline
\end{array}
\hspace{+1.3cm}
$$

\newpage

\underline{\large {\bf TABLE}\ V:}

\vspace{1cm}

\framebox{${\bf d=4}$}

\bigskip

$$
\begin{array}{|c|c|c|}
\hline &&\\
& {\rm Radiated\ momentum\ flux}
&{\rm Particular\ Projections} \\
&&\\ \hline &&\\
{\rm rank\ 0\ tensor}& {d{\cal P}_{4|0}^\mu\over d\tau} =
{4\pi\over 3}\dot u^2 u^\mu &
P_0^{(4|0)} = I_{4|0}=-\frac{4\pi}{3}\kappa_1^2 
\\ &&\\
{\rm vector} &{d{\cal P}_{4|1}^\mu\over d\tau} = -{8\pi\over 3}\dot
u^2 u^\mu
&  P_0^{(4|1)} =I_{4|1} =\frac{8\pi}{3}\kappa_1^2 
\\ &&\\
{\rm rank}\ 2\ {\rm tensor}& {d{\cal P}_{4|2}^\mu\over d\tau} =
-{20\pi\over 3}\dot u^2 u^\mu & P_0^{(4|2)} =I_{4|2}
=\frac{20\pi}{3}\kappa_1^2
\\ &&\\
{\rm gravity} &{d{\cal P}_{4|g}^\mu\over d\tau}= {\cal
P}_{4|2}^\mu-{1\over 2} {d{\cal P}_{4|0}^\mu\over d\tau}=
-{22\pi\over 3}\dot u^2 u^\mu & P_0^{(4|g)} =I_{4|g} =
I_{4|2}=\frac{22\pi}{3}\kappa_1^2
\\ &&\\
{\rm rank}\ 3\ {\rm tensor} &{d{\cal P}_{4|3}^\mu\over d\tau} =
-{32\pi\over 3}\dot u^2 u^\mu
& P_0^{(4|3)}=I_{4|3} = \frac{32\pi}{3}\kappa_1^2 
\\&&\\\hline&&\\
{\rm rank}\ s\ {\rm tensor}&{d{\cal P}^\mu_{4|s}\over d\tau}=
{4-12s\over 3}\pi\dot u^2 u^\mu& P_0^{(4|s)}=I_{4|s} =
\frac{12s-4}{3}\pi\kappa_1^2\\ &&\\ \hline
\end{array}
$$

\newpage

\framebox{${\bf d=6}$}

\bigskip

$$
\begin{array}{|c|c|}
\hline &\\
 {\rm Radiated\ momentum\ flux}
&{\rm Particular\ Projections} \\\hline &\\&\\
 {d{\cal P}_{6|0}^\mu\over d\tau} = -{8\pi^2\over 3}\dot u^4
u^\mu+{8\pi^2\over 15}\ddot u^2 u^\mu+{32\pi^2\over 35}(\dot u\ddot
u)\dot u^\mu-{64\pi^2\over 105}\dot u^2\ddot u^\mu & I_{6|0}
=\frac{8\pi^2}{105} \Big(20\kappa_1^4 + 7\dot\kappa_1^2 +
7\kappa_1^2\kappa_2^2\Big)\\
&\\
&P^{(6|0)}_0=-{8\pi^2\over 15}(4\kappa_1^4+\kappa_1^2\kappa_2^2)\\
&\\
{d{\cal P}_{6|1}^\mu\over d\tau} = {16\pi^2\over 3}\dot u^4
u^\mu-{32\pi^2\over 15}\ddot u^2 u^\mu-{16\pi^2\over 7}(\dot u\ddot
u)\dot u^\mu+{16\pi^2\over 35}\dot u^2\ddot u^\mu &I_{6|1} =
\frac{32\pi^2}{105}\Big(9\kappa_1^4 + 7\dot\kappa_1^2
+ 7\kappa_1^2\kappa_2^2\Big) \\ &\\
&P_0^{(6|1)} = {16\pi^2\over
15}(3\kappa_1^4+2\kappa_1^2\kappa_2^2)\\&\\  {d{\cal
P}_{6|2}^\mu\over d\tau} = {16\pi^2\over 3}\dot u^4
u^\mu-{24\pi^2\over 5}\ddot u^2 u^\mu-{192\pi^2\over 35}(\dot u\ddot
u)\dot u^\mu+{32\pi^2\over 21}\dot u^2\ddot u^\mu & I_{6|2} =
\frac{8\pi^2}{105}\Big(-13\kappa_1^4 +63\dot\kappa_1^2 +
63\kappa_1^2\kappa_2^2\Big)
\\ &\\&P_0^{(6|2)}={8\pi^2\over 15}(\kappa_1^4+
9\kappa_1^2\kappa_2^2)\\&\\ {d{\cal P}_{6|g}^\mu\over d\tau} =
{20\pi^2\over 3}\dot u^4 u^\mu-{76\pi^2\over 15}\ddot u^2
u^\mu-{208\pi^2\over 35}(\dot u\ddot u)\dot u^\mu+{64\over 35}\dot
u^2\ddot u^\mu &I_{6|g} = I_{6|2}-\frac{1}{2}I_{6|0} =\\&\\&=
\frac{4\pi^2}{105}\Big(-46\kappa_1^4 + 119\dot\kappa_1^2
+ 119\kappa_1^2\kappa_2^2\Big) \\
&\\&P_0^{(6|g)}={4\pi^2\over
15}(6\kappa_1^4+19\kappa_1^2\kappa_2^2)\\&\\  {d{\cal
P}_{6|3}^\mu\over d\tau} =-{8\pi^2\over 3}\dot u^4
u^\mu-{112\pi^2\over 15}\ddot u^2 u^\mu-{304\pi^2\over 35}(\dot
u\ddot u)\dot u^\mu+{272\pi^2\over 105}\dot u^2\ddot u^\mu &I_{6|3}
= \frac{8\pi^2}{105}\Big(-167\kappa_1^4 + 98\dot\kappa_1^2
+ 98\kappa_1^2\kappa_2^2\Big) \\
&\\&P_0^{(6|3)}=-{8\pi^2\over
15}(19\kappa_1^4-14\kappa_1^2\kappa_2^2)\\&\\\hline&\\{d{\cal
P}_{6|s}^\mu\over d\tau} =\pi^2\left[{19\over 3}-(2s-3)^2\right]\dot
u^4 u^\mu+{8\pi^2\over 15}(1-5s)\ddot u^2
u^\mu+&I_{6|s}=-\pi^2\left(4s^2-{124\over 15}s+{32\over
21}\right)\kappa_1^4 +\\&\\+{16\pi^2\over 35}(2-7s)(\dot u\ddot
u)\dot u^\mu+{16\pi^2\over 105}(7s-4)\dot u^2\ddot u^\mu &+
{8\pi^2\over 15}(5s-1)\dot\kappa_1^2 + {8\pi^2\over
15}(5s-1)\kappa_1^2\kappa_2^2\Big)\\&\\&P_0^{(6|s)}=
-\pi^2\left(4s^2-{28\over 3}s+{32\over 15}\right)4\kappa_1^4+\\&\\&+
{8\pi^2\over 15}(5s-1)\kappa_1^2\kappa_2^2)\\&\\
\hline
\end{array}
$$

\newpage

\framebox{${\bf d=8}$}

\bigskip

$$\hspace{-1.3cm}
\begin{array}{|c|c|}
\hline&\\
{d{\cal P}_{8|0}^\mu\over d\tau} =\left\{{80\pi^3\over 9}\dot
u^6-{1312\pi^3\over 189}(\dot u\ddot u)^2-{320\pi^3\over 189}\dot
u^2\ddot u^2+{352\pi^3\over 189}\dot u^2(\dot u\dot{\ddot
u})+{16\pi^3\over 105}\dot{\ddot
u}^2\right\}u^\mu-&P_0^{(8|0)}=-{16\pi^3\over
315}(3\kappa_1^2\kappa_2^2\kappa_3^2+108\kappa_1^6+\\-\left\{{9056\pi^3\over
2079}\dot u^2(\dot u\ddot u)-{64\pi^3\over 189}\ddot u\dot{\ddot
u}\right\}\dot u^\mu+\left\{{4160\pi^3\over 2079}\dot
u^4+{64\pi^3\over 189}\dot u\dot{\ddot u}\right\}\ddot
u^\mu-{544\pi^3\over 945}(\dot u\ddot u)\dot{\ddot
u}^\mu&+64\kappa_1^4\kappa_2^2+3\kappa_1^2\kappa_2^4)\\
 {d{\cal P}_{8|1}^\mu\over d\tau} =\left\{-{160\pi^3\over 9}\dot
u^6+{12448\pi^3\over 945}(\dot u\ddot u)^2+{5888\pi^3\over 945}\dot
u^2\ddot u^2-{4864\pi^3\over 945}\dot u^2(\dot u\dot{\ddot
u})-{32\pi^3\over 35}\dot{\ddot
u}^2\right\}u^\mu+&P_0^{(8|1)}={32\pi^3\over
315}(9\kappa_1^2\kappa_2^2\kappa_3^2+72\kappa_1^6+\\+\left\{{78272\pi^3\over
10395}\dot u^2(\dot u\ddot u)-{1408\pi^3\over 945}\ddot u\dot{\ddot
u}\right\}\dot u^\mu-\left\{{29888\pi^3\over 10395}\dot
u^4+{832\pi^3\over 945}\dot u\dot{\ddot u}\right\}\ddot
u^\mu+{64\pi^3\over 189}(\dot u\ddot
u)\dot{\ddot u}^\mu & +94\kappa_1^4\kappa_2^2+9\kappa_1^2\kappa_2^4)\\ &\\
{d{\cal P}_{8|2}^\mu\over d\tau} =\left\{-{5552\pi^3\over 315}\dot
u^6+{11296\pi^3\over 945}(\dot u\ddot u)^2+{2272\pi^3\over 189}\dot
u^2\ddot u^2-{7456\pi^3\over 945}\dot u^2(\dot u\dot{\ddot
u})-{208\pi^3\over 105}\dot{\ddot
u}^2\right\}u^\mu+&P_0^{(8|2)}={16\pi^3\over
315}(39\kappa_1^2\kappa_2^2\kappa_3^2-6\kappa_1^6+\\+\left\{{87776\pi^3\over
10395}\dot u^2(\dot u\ddot u)-{448\pi^3\over 135}\ddot u\dot{\ddot
u}\right\}\dot u^\mu-\left\{{55232\pi^3\over 10395}\dot
u^4+{1984\pi^3\over 945}\dot u\dot{\ddot u}\right\}\ddot
u^\mu+{1184\pi^3\over 945}(\dot u\ddot u)\dot{\ddot u}^\mu &
+314\kappa_1^4\kappa_2^2+39\kappa_1^2\kappa_2^4)  \\
&\\
{d{\cal P}_{8|g}^\mu\over d\tau} =\left\{-{6952\pi^3\over 315}\dot
u^6+{14576\pi^3\over 945}(\dot u\ddot u)^2+{2432\pi^3\over 189}\dot
u^2\ddot u^2-{8336\pi^3\over 945}\dot u^2(\dot u\dot{\ddot
u})-{72\pi^3\over 35}\dot{\ddot
u}^2\right\}u^\mu+&P_0^{(8|g)}={72\pi^3\over
35}\kappa_1^2\kappa_2^2\kappa_3^2+{256\pi^3\over
105}\kappa_1^6+\\+\left\{{110416\pi^3\over 10395}\dot u^2(\dot
u\ddot u)-{3296\pi^3\over 945}\ddot u\dot{\ddot u}\right\}\dot
u^\mu-\left\{{9376\pi^3\over 1485}\dot u^4+{2144\pi^3\over 945}\dot
u\dot{\ddot u}\right\}\ddot u^\mu+{208\pi^3\over 135}(\dot u\ddot
u)\dot{\ddot u}^\mu & +{5536\pi^3\over
315}\kappa_1^4\kappa_2^2+{72\pi^3\over
35}\kappa_1^2\kappa_2^4 \\ &\\
{d{\cal P}_{8|3}^\mu\over d\tau} =\left\{{592\pi^3\over 315}\dot
u^6-{10016\pi^3\over 945}(\dot u\ddot u)^2+{14816\pi^3\over 945}\dot
u^2\ddot u^2-{6016\pi^3\over 945}\dot u^2(\dot u\dot{\ddot
u})-{64\pi^3\over 21}\dot{\ddot
u}^2\right\}u^\mu-&P_0^{(8|3)}={16\pi^3\over
315}(60\kappa_1^2\kappa_2^2\kappa_3^2-411\kappa_1^6+\\-\left\{{16768\pi^3\over
10395}\dot u^2(\dot u\ddot u)+{4864\pi^3\over 945}\ddot u\dot{\ddot
u}\right\}\dot u^\mu-\left\{{55232\pi^3\over 10395}\dot
u^4+{448\pi^3\over 135}\dot u\dot{\ddot u}\right\}\ddot
u^\mu+{2048\pi^3\over 945}(\dot u\ddot u)\dot{\ddot u}^\mu &
+314\kappa_1^4\kappa_2^2+60\kappa_1^2\kappa_2^4)\\&\\\hline&\\
{d{\cal P}_{8|s}^\mu\over d\tau} =\left\{-{16\pi^3\over
9}\left({7s^3\over 10}-{135\over 14}s^2+{838\over 35}s-5\right)\dot
u^6+{32\pi^3\over 3}\left(-s^2+{101\over 35}s-{41\over
63}\right)(\dot u\ddot u)^2+\right.&P_0^{(8|s)}={16\pi^3\over
105}(7s-1)(\kappa_1^2\kappa_2^2\kappa_3^2+\kappa_1^2\kappa_2^4)+
\\\left.+{16\pi^3\over 3}\left(-{s^2\over
5}+{59\over 35}s-{20\over 63}\right)\dot u^2\ddot
u^2+32\pi^3\left({s^2\over 15}-{2s\over 7}+{11\over 189}\right)\dot
u^2(\dot u\dot{\ddot u})+\right.&+{8\pi^3\over
315}(49s^3-549s^2+1004s-216)\kappa_1^6-\\\left.+{16\pi^3\over
15}\left({1\over 7}-s\right)\dot{\ddot
u}^2\right\}u^\mu+\left\{{32\pi^3\over 7}\left(-{6s^2\over
5}+{19\over 5}s-{283\over 297}\right)\dot u^2(\dot u\ddot
u)+{64\pi^3\over 7}\left({1\over 27}-{s\over 5}\right)\ddot
u\dot{\ddot u}\right\}\dot u^\mu+&-16\pi^3\left({s^2\over
5}-s+{64\over 315}\right)\kappa_1^4\kappa_2^2
\\+\left\{{64\pi^3\over
21}\left({2s^2\over 5}-2s+{65\over 99}\right)\dot u^4+{64\pi^3\over
21}\left({1\over 9}-{2s\over 5}\right)\dot u\dot{\ddot
u}\right\}\ddot u^\mu+{32\pi^3\over 35}\left(s-{17\over
27}\right)(\dot u\ddot
u)\dot{\ddot u}^\mu &\\
&\\ \hline
\end{array}
$$

\bigskip

$$
I_{8|0}=-{16\pi^3\over 1155}\left[276\kappa_1^6+{1892\over
9}\kappa_1^4\kappa_2^2+ {1199\over
3}\kappa_1^2\dot\kappa_1^2-88\kappa_1^3\ddot\kappa_1+11\left\{\kappa_1^2
\kappa_2^2\kappa_3^2+(\kappa_1\kappa_2^2- \ddot\kappa_1)^2+
(2\kappa_2\dot\kappa_1+ \kappa_1\dot\kappa_2)^2\right\}\right]
$$


$$
I_{8|1}={32\pi^3\over 385}\left[64\kappa_1^6+{2816\over
27}\kappa_1^4\kappa_2^2+ {1100\over
9}\kappa_1^2\dot\kappa_1^2-{88\over
3}\kappa_1^3\ddot\kappa_1+11\left\{\kappa_1^2
\kappa_2^2\kappa_3^2+(\kappa_1\kappa_2^2- \ddot\kappa_1)^2+
(2\kappa_2\dot\kappa_1+ \kappa_1\dot\kappa_2)^2\right\}\right]
$$


$$
I_{8|2}={16\pi^3\over 1155}\left[-254\kappa_1^6+{8998\over
9}\kappa_1^4\kappa_2^2+ {517\over
3}\kappa_1^2\dot\kappa_1^2-132\kappa_1^3\ddot\kappa_1+143\left\{\kappa_1^2
\kappa_2^2\kappa_3^2+(\kappa_1\kappa_2^2- \ddot\kappa_1)^2+
(2\kappa_2\dot\kappa_1+ \kappa_1\dot\kappa_2)^2\right\}\right]
$$


$$
I_{8|g}={8\pi^3\over 1155}\left[-232\kappa_1^6+{19888\over
9}\kappa_1^4\kappa_2^2+ {2233\over
3}\kappa_1^2\dot\kappa_1^2-352\kappa_1^3\ddot\kappa_1+297\left\{\kappa_1^2
\kappa_2^2\kappa_3^2+(\kappa_1\kappa_2^2- \ddot\kappa_1)^2+
(2\kappa_2\dot\kappa_1+ \kappa_1\dot\kappa_2)^2\right\}\right]
$$


$$
I_{8|3}={16\pi^3\over 1155}\left[-1651\kappa_1^6+{8206\over
9}\kappa_1^4\kappa_2^2- {6248\over
3}\kappa_1^2\dot\kappa_1^2+220\kappa_1^3\ddot\kappa_1+220\left\{\kappa_1^2
\kappa_2^2\kappa_3^2+(\kappa_1\kappa_2^2- \ddot\kappa_1)^2+
(2\kappa_2\dot\kappa_1+ \kappa_1\dot\kappa_2)^2\right\}\right]
$$

\bigskip

\hrule

\bigskip

$$
I_{8|s}={16\pi^3\over 105}\left[\left({49\over 6}s^3-{167\over
2}s^2+{406\over 3}s-{276\over 11}\right)\kappa_1^6-
\left(21s^2-97s+{172\over 9}\right)\kappa_1^4\kappa_2^2-
\left(77s^2-180s+{109\over 3}\right)\kappa_1^2\dot\kappa_1^2+
\right.
$$
$$
+\left. 2(7s^2-9s+4)\kappa_1^3\ddot\kappa_1+(7s-1)\left\{\kappa_1^2
\kappa_2^2\kappa_3^2+(\kappa_1\kappa_2^2- \ddot\kappa_1)^2+
(2\kappa_2\dot\kappa_1+ \kappa_1\dot\kappa_2)^2\right\}\right]
$$

\newpage

\framebox{${\bf d=10}$}

\bigskip

$$
{d{\cal P}_{10|0}^\mu\over d\tau} =\left\{-{1120\pi^4\over 27} \dot
u^8-{1088\pi^4\over 315}(\dot u\dot{\ddot u})^2-{8192\pi^4\over
3465}{\ddot u}^4 +{32\pi^4\over 945}\ddot{\ddot
u}^2+{28480\pi^4\over 429}\dot u^2(\dot u\ddot u)^2+{5344\pi^4\over
1287}\dot u^4\ddot u^2 -\right.
$$

$$
-\left. {8896\pi^4\over 429}\dot u^4(\dot u\dot {\ddot
u})-{2176\pi^4\over 495}\ddot u^2(\dot u\dot{\ddot u})
 -{1280\pi^4\over 693}(\dot u\ddot u)(\ddot u\dot{\ddot
u})-{160\pi^4\over 231}\dot u^2\dot{\ddot u}^2 +{164\pi^4\over
77}\dot u^2(\ddot u\ddot{\ddot u})+{128\pi^4\over 63}(\dot u\ddot
u)(\dot u\ddot{\ddot u})\right\}u^\mu+
$$

$$
+ \left\{{64\pi^4\over 693}(\dot {\ddot u}\ddot{\ddot u})-
{38656\pi^4\over 9009}(\dot u\ddot u)(\dot u\dot{\ddot
u})+{896\pi^4\over 1287}\dot u^2(\dot u\ddot{\ddot
u})+{2880\pi^4\over 143}\dot u^4(\dot u\ddot u)-{320\pi^4\over
819}\dot u^2(\ddot u\dot{\ddot u})-{18944\pi^4\over 3003}\ddot
u^2(\dot u\ddot u)\right\}\dot u^\mu-
$$

$$
-\left\{{37760\pi^4\over 3861}\dot u^6+{2560\pi^4\over 1001}(\dot
u{\ddot u})^2+{256\pi^4\over 2079}(\ddot u\ddot {\ddot
u})-{33472\pi^4\over 9009}\dot u^2(\dot u\dot{\ddot
u})-{2624\pi^4\over 9009}\dot u^2\ddot u^2\right\}\ddot u^\mu+
$$

$$
+\left\{{64\pi^4\over 693}\dot u\ddot{\ddot u}+{33920\pi^4\over
9009}\dot u^2(\dot u\ddot u)\right\}\dot{\ddot
u}^\mu-\left\{{1472\pi^4\over 10395}\ddot u^2-{1856\pi^4\over
10395}\dot u\dot{\ddot u}-{2176\pi^4\over 9009}\dot
u^4\right\}\ddot{\ddot u}^\mu
$$

\bigskip

\hrule

\bigskip

$$
{d{\cal P}_{10|1}^\mu\over d\tau} =\left\{{2240\pi^4\over 27} \dot
u^8+{704\pi^4\over 105}(\dot u\dot{\ddot u})^2+{7552\pi^4\over
1485}{\ddot u}^4 -{256\pi^4\over 945}\ddot{\ddot
u}^2-{15680\pi^4\over 117}\dot u^2(\dot u\ddot u)^2-{320\pi^4\over
13}\dot u^4\ddot u^2 +\right.
$$

$$
+\left. {5440\pi^4\over 17}\dot u^4(\dot u\dot {\ddot
u})+{63424\pi^4\over 10395}\ddot u^2(\dot u\dot{\ddot u})
 +{640\pi^4\over 77}(\dot u\ddot u)(\ddot u\dot{\ddot
u})+{7360\pi^4\over 2079}\dot u^2\dot{\ddot u}^2 -{7424\pi^4\over
2079}\dot u^2(\ddot u\ddot{\ddot u})-{128\pi^4\over 21}(\dot u\ddot
u)(\dot u\ddot{\ddot u})\right\}u^\mu-
$$

$$
- \left\{{1216\pi^4\over 2079}(\dot {\ddot u}\ddot{\ddot u})+
{4352\pi^4\over 819}(\dot u\ddot u)(\dot u\dot{\ddot
u})-{1984\pi^4\over 1001}\dot u^2(\dot u\ddot{\ddot
u})-{50560\pi^4\over 1287}\dot u^4(\dot u\ddot u)+{8000\pi^4\over
3861}\dot u^2(\ddot u\dot{\ddot u})+{112064\pi^4\over 9009}\ddot
u^2(\dot u\ddot u)\right\}\dot u^\mu+
$$

$$
+\left\{{59200\pi^4\over 3861}\dot u^6-{3200\pi^4\over 819}(\dot
u{\ddot u})^2-{128\pi^4\over 231}(\ddot u\ddot {\ddot
u})+{58048\pi^4\over 9009}\dot u^2(\dot u\dot{\ddot
u})-{53632\pi^4\over 27027}\dot u^2\ddot u^2\right\}\ddot u^\mu-
$$

$$
-\left\{{512\pi^4\over 2079}\dot u\ddot{\ddot u}-{139520\pi^4\over
27027}\dot u^2(\dot u\ddot u)\right\}\dot{\ddot
u}^\mu+\left\{{128\pi^4\over 2079}\ddot u^2+{64\pi^4\over 693}\dot
u\dot{\ddot u}+{4288\pi^4\over 27027}\dot u^4\right\}\ddot{\ddot
u}^\mu
$$

\bigskip

\hrule

\bigskip

$$
{d{\cal P}_{10|2}^\mu\over d\tau} =\left\{{160000\pi^4\over 2079}
\dot u^8+{2048\pi^4\over 315}(\dot u\dot{\ddot
u})^2+{10016\pi^4\over 1485}{\ddot u}^4 -{544\pi^4\over
945}\ddot{\ddot u}^2-{1429504\pi^4\over 9009}\dot u^2(\dot u\ddot
u)^2-{1424480\pi^4\over 27027}\dot u^4\ddot u^2 +\right.
$$

$$
+\left. {210304\pi^4\over 3861}\dot u^4(\dot u\dot {\ddot
u})+{4096\pi^4\over 2079}\ddot u^2(\dot u\dot{\ddot u})
 +{59776\pi^4\over 3465}(\dot u\ddot u)(\ddot u\dot{\ddot
u})+{74464\pi^4\over 10395}\dot u^2\dot{\ddot u}^2 -{70208\pi^4\over
10395}\dot u^2(\ddot u\ddot{\ddot u})-{3328\pi^4\over 315}(\dot
u\ddot u)(\dot u\ddot{\ddot u})\right\}u^\mu-
$$

$$
- \left\{{2624\pi^4\over 2079}(\dot {\ddot u}\ddot{\ddot u})+
{37120\pi^4\over 27027}(\dot u\ddot u)(\dot u\dot{\ddot
u})-{89344\pi^4\over 27027}\dot u^2(\dot u\ddot{\ddot
u})-{469120\pi^4\over 9009}\dot u^4(\dot u\ddot u)+{122560\pi^4\over
27027}\dot u^2(\ddot u\dot{\ddot u})+{476800\pi^4\over 27027}\ddot
u^2(\dot u\ddot u)\right\}\dot u^\mu+
$$

$$
+\left\{{49984\pi^4\over 2457}\dot u^6-{188800\pi^4\over 27027}(\dot
u{\ddot u})^2-{2560\pi^4\over 2079}(\ddot u\ddot {\ddot
u})+{37952\pi^4\over 3861}\dot u^2(\dot u\dot{\ddot
u})-{57536\pi^4\over 9009}\dot u^2\ddot u^2\right\}\ddot u^\mu-
$$

$$
-\left\{{1216\pi^4\over 2079}\dot u\ddot{\ddot u}-{11392\pi^4\over
1001}\dot u^2(\dot u\ddot u)\right\}\dot{\ddot
u}^\mu+\left\{{2752\pi^4\over 10395}\ddot u^2+{3776\pi^4\over
10395}\dot u\dot{\ddot u}+{66368\pi^4\over 135135}\dot
u^4\right\}\ddot{\ddot u}^\mu
$$

\bigskip

\hrule

\bigskip

$$
{d{\cal P}_{10|g}^\mu\over d\tau} =\left\{{203120\pi^4\over 2079}
\dot u^8+{288\pi^4\over 35}(\dot u\dot{\ddot u})^2+{16480\pi^4\over
2079}{\ddot u}^4 -{16\pi^4\over 27}\ddot{\ddot
u}^2-{1728544\pi^4\over 9009}\dot u^2(\dot u\ddot
u)^2-{1480592\pi^4\over 27027}\dot u^4\ddot u^2 +\right.
$$

$$
+\left. {250336\pi^4\over 3861}\dot u^4(\dot u\dot {\ddot
u})+{43328\pi^4\over 10395}\ddot u^2(\dot u\dot{\ddot u})
 +{20992\pi^4\over 1155}(\dot u\ddot u)(\ddot u\dot{\ddot
u})+{11152\pi^4\over 1485}\dot u^2\dot{\ddot u}^2 -{74528\pi^4\over
10395}\dot u^2(\ddot u\ddot{\ddot u})-{1216\pi^4\over 105}(\dot
u\ddot u)(\dot u\ddot{\ddot u})\right\}u^\mu-
$$

$$
- \left\{{2720\pi^4\over 2079}(\dot {\ddot u}\ddot{\ddot u})+
{95104\pi^4\over 27027}(\dot u\ddot u)(\dot u\dot{\ddot
u})-{98752\pi^4\over 27027}\dot u^2(\dot u\ddot{\ddot
u})-{559840\pi^4\over 9009}\dot u^4(\dot u\ddot u)+{127840\pi^4\over
27027}\dot u^2(\ddot u\dot{\ddot u})+{562048\pi^4\over 27027}\ddot
u^2(\dot u\ddot u)\right\}\dot u^\mu+
$$

$$
+\left\{{75776\pi^4\over 3003}\dot u^6-{223360\pi^4\over 27027}(\dot
u{\ddot u})^2-{128\pi^4\over 99}(\ddot u\ddot {\ddot
u})+{315872\pi^4\over 27027}\dot u^2(\dot u\dot{\ddot
u})-{8032\pi^4\over 1287}\dot u^2\ddot u^2\right\}\ddot u^\mu-
$$

$$
-\left\{{1312\pi^4\over 2079}\dot u\ddot{\ddot u}-{119488\pi^4\over
9009}\dot u^2(\dot u\ddot u)\right\}\dot{\ddot
u}^\mu+\left\{{3488\pi^4\over 10395}\ddot u^2+{224\pi^4\over
495}\dot u\dot{\ddot u}+{82688\pi^4\over 135135}\dot
u^4\right\}\ddot{\ddot u}^\mu
$$

\bigskip

\hrule

\bigskip

$$
{d{\cal P}_{10|3}^\mu\over d\tau} =\left\{-{3104\pi^4\over 231} \dot
u^8-{256\pi^4\over 63}(\dot u\dot{\ddot u})^2+{9056\pi^4\over
3465}{\ddot u}^4 -{832\pi^4\over 945}\ddot{\ddot
u}^2-{105152\pi^4\over 3003}\dot u^2(\dot u\ddot
u)^2-{3485408\pi^4\over 45045}\dot u^4\ddot u^2 +\right.
$$

$$
+\left. {637568\pi^4\over 45045}\dot u^4(\dot u\dot {\ddot
u})-{6464\pi^4\over 385}\ddot u^2(\dot u\dot{\ddot u})
 +{86528\pi^4\over 3465}(\dot u\ddot u)(\ddot u\dot{\ddot
u})+{35264\pi^4\over 3465}\dot u^2\dot{\ddot u}^2 -{30208\pi^4\over
3465}\dot u^2(\ddot u\ddot{\ddot u})-{512\pi^4\over 45}(\dot u\ddot
u)(\dot u\ddot{\ddot u})\right\}u^\mu-
$$

$$
- \left\{{64\pi^4\over 33}(\dot {\ddot u}\ddot{\ddot u})-
{2304\pi^4\over 143}(\dot u\ddot u)(\dot u\dot{\ddot
u})-{29504\pi^4\over 9009}\dot u^2(\dot u\ddot{\ddot
u})-{237376\pi^4\over 9009}\dot u^4(\dot u\ddot u)+{63040\pi^4\over
9009}\dot u^2(\ddot u\dot{\ddot u})+{1088\pi^4\over 117}\ddot
u^2(\dot u\ddot u)\right\}\dot u^\mu+
$$

$$
+\left\{{35968\pi^4\over 3861}\dot u^6-{60160\pi^4\over 9009}(\dot
u{\ddot u})^2-{3968\pi^4\over 2079}(\ddot u\ddot {\ddot
u})+{58048\pi^4\over 9009}\dot u^2(\dot u\dot{\ddot
u})-{121600\pi^4\over 9009}\dot u^2\ddot u^2\right\}\ddot u^\mu-
$$

$$
-\left\{{640\pi^4\over 693}\dot u\ddot{\ddot u}-{134144\pi^4\over
9009}\dot u^2(\dot u\ddot u)\right\}\dot{\ddot
u}^\mu+\left\{{4864\pi^4\over 10395}\ddot u^2+{6592\pi^4\over
10395}\dot u\dot{\ddot u}+{4864\pi^4\over 6435}\dot
u^4\right\}\ddot{\ddot u}^\mu
$$

\bigskip

\hrule

\vspace{-0.10cm}

\hrule

\bigskip

$$
P_0^{(10|0)}=-{32\pi^4\over
135135}\left[82368\kappa_1^8+80652\kappa_1^6\kappa_2^2+
12012\kappa_1^4\kappa_2^4+6149\kappa_1^4\kappa_2^2\kappa_3^2+
143(\kappa_1^2\kappa_2^6+\kappa_1^2\kappa_2^2\kappa_3^2\kappa_4^2+
2\kappa_1^2\kappa_2^4\kappa_3^2+\kappa_1^2\kappa_2^2\kappa_3^4)\right]
$$

$$
P_0^{(10|1)}={64\pi^4\over
135135}\left[51480\kappa_1^8+97812\kappa_1^6\kappa_2^2+
25311\kappa_1^4\kappa_2^4+13871\kappa_1^4\kappa_2^2\kappa_3^2+
572(\kappa_1^2\kappa_2^6+\kappa_1^2\kappa_2^2\kappa_3^2\kappa_4^2+
2\kappa_1^2\kappa_2^4\kappa_3^2+\kappa_1^2\kappa_2^2\kappa_3^4)\right]
$$

$$\hspace{-0.8cm}
P_0^{(10|2)}={32\pi^4\over
135135}\left[-23634\kappa_1^8+247104\kappa_1^6\kappa_2^2+
99099\kappa_1^4\kappa_2^4+53911\kappa_1^4\kappa_2^2\kappa_3^2+
2431(\kappa_1^2\kappa_2^6+\kappa_1^2\kappa_2^2\kappa_3^2\kappa_4^2+
2\kappa_1^2\kappa_2^4\kappa_3^2+\kappa_1^2\kappa_2^2\kappa_3^4)\right]
$$

$$\hspace{-0.8cm}
P_0^{(10|g)}={16\pi^4\over
135135}\left[35100\kappa_1^8+574860\kappa_1^6\kappa_2^2+
210210\kappa_1^4\kappa_2^4+113971\kappa_1^4\kappa_2^2\kappa_3^2+
5005(\kappa_1^2\kappa_2^6+\kappa_1^2\kappa_2^2\kappa_3^2\kappa_4^2+
2\kappa_1^2\kappa_2^4\kappa_3^2+\kappa_1^2\kappa_2^2\kappa_3^4)\right]
$$

$$\hspace{-1.cm}
P_0^{(10|3)}={32\pi^4\over
135135}\left[-302419\kappa_1^8+108537\kappa_1^6\kappa_2^2+
133419\kappa_1^4\kappa_2^4+72358\kappa_1^4\kappa_2^2\kappa_3^2+
3718(\kappa_1^2\kappa_2^6+\kappa_1^2\kappa_2^2\kappa_3^2\kappa_4^2+
2\kappa_1^2\kappa_2^4\kappa_3^2+\kappa_1^2\kappa_2^2\kappa_3^4)\right]
$$

\bigskip

\hrule

\vspace{-0.10cm}

\hrule

\bigskip

$$\hspace{-.5cm}
I_{10|0}=-{32\pi^4\over
135135}\left(66524\kappa_1^6\kappa_2^2+150140\kappa_1^4\dot\kappa_1^2+11336
\kappa_1^4\kappa_2^4-2704\kappa_1^4\kappa_2\ddot\kappa_2-
22048\kappa_1^3\ddot\kappa_1\kappa_2^2+18278\kappa_1\ddot\kappa_1\dot\kappa_1^2
-8684\kappa_1^2\dot\kappa_1\dot{\ddot\kappa}_1+\right.
$$

$$+
6903\dot\kappa_1^4+
5629\kappa_1^4\kappa_2^2\kappa_3^2+55872\kappa_1^8-41904\kappa_1^5\ddot\kappa_1+
12207\kappa_1^2\ddot\kappa_1^2+37440\kappa_1^3\dot\kappa_1\kappa_2\dot\kappa_2+
2925\kappa_1^4\dot\kappa_2^2+48880\kappa_1^2\dot\kappa_1^2\kappa_2^2+
$$

$$
+143[(\dot{\ddot\kappa}_1-3\dot\kappa_1\kappa_2^2-
3\kappa_1\kappa_2\dot\kappa_2)^2+(3\ddot\kappa_1\kappa_2+3\dot\kappa_1\dot\kappa_2+
\kappa_1\ddot\kappa_2-\kappa_1\kappa_2^3)^2+(2\kappa_1\dot\kappa_2\kappa_3+
\kappa_1\kappa_2\dot\kappa_3+3\dot\kappa_1\kappa_2\kappa_3)^2+
$$

$$\left.+
\kappa_1^2\kappa_2^2\kappa_3^2(2\kappa_2^2+\kappa_3^2+\kappa_4^2)-
2\kappa_1\kappa_2\kappa_3^2(3\ddot\kappa_1\kappa_2+\kappa_1\ddot\kappa_2+
3\dot\kappa_1\dot\kappa_2)]\right)
$$

\bigskip

\hrule

\bigskip

$$\hspace{-.5cm}
I_{10|1}={64\pi^4\over
135135}\left(82152\kappa_1^6\kappa_2^2+109712\kappa_1^4\dot\kappa_1^2+24076
\kappa_1^4\kappa_2^4-5226\kappa_1^4\kappa_2\ddot\kappa_2-
41028\kappa_1^3\ddot\kappa_1\kappa_2^2+15600\kappa_1\ddot\kappa_1\dot\kappa_1^2
-10816\kappa_1^2\dot\kappa_1\dot{\ddot\kappa}_1+\right.
$$

$$+
5200\dot\kappa_1^4+
12701\kappa_1^4\kappa_2^2\kappa_3^2+36000\kappa_1^8-32400\kappa_1^5\ddot\kappa_1+
11700\kappa_1^2\ddot\kappa_1^2+64220\kappa_1^3\dot\kappa_1\kappa_2\dot\kappa_2+
7475\kappa_1^4\dot\kappa_2^2+92248\kappa_1^2\dot\kappa_1^2\kappa_2^2+
$$

$$
+572[(\dot{\ddot\kappa}_1-3\dot\kappa_1\kappa_2^2-
3\kappa_1\kappa_2\dot\kappa_2)^2+(3\ddot\kappa_1\kappa_2+3\dot\kappa_1\dot\kappa_2+
\kappa_1\ddot\kappa_2-\kappa_1\kappa_2^3)^2+(2\kappa_1\dot\kappa_2\kappa_3+
\kappa_1\kappa_2\dot\kappa_3+3\dot\kappa_1\kappa_2\kappa_3)^2+
$$

$$\left.+
\kappa_1^2\kappa_2^2\kappa_3^2(2\kappa_2^2+\kappa_3^2+\kappa_4^2)-
2\kappa_1\kappa_2\kappa_3^2(3\ddot\kappa_1\kappa_2+\kappa_1\ddot\kappa_2+
3\dot\kappa_1\dot\kappa_2)]\right)
$$

\bigskip

\hrule

\bigskip

$$\hspace{-.5cm}
I_{10|2}={32\pi^4\over
135135}\left(187782\kappa_1^6\kappa_2^2+8980\kappa_1^4\dot\kappa_1^2+93483
\kappa_1^4\kappa_2^4-18460\kappa_1^4\kappa_2\ddot\kappa_2-
139828\kappa_1^3\ddot\kappa_1\kappa_2^2+13754\kappa_1\ddot\kappa_1\dot\kappa_1^2
-31356\kappa_1^2\dot\kappa_1\dot{\ddot\kappa}_1+\right.
$$

$$+
3250\dot\kappa_1^4+
48711\kappa_1^4\kappa_2^2\kappa_3^2-44606\kappa_1^8-2478\kappa_1^5\ddot\kappa_1+
12675\kappa_1^2\ddot\kappa_1^2+232544\kappa_1^3\dot\kappa_1\kappa_2\dot\kappa_2+
30251\kappa_1^4\dot\kappa_2^2+353522\kappa_1^2\dot\kappa_1^2\kappa_2^2+
$$

$$
+2431[(\dot{\ddot\kappa}_1-3\dot\kappa_1\kappa_2^2-
3\kappa_1\kappa_2\dot\kappa_2)^2+(3\ddot\kappa_1\kappa_2+3\dot\kappa_1\dot\kappa_2+
\kappa_1\ddot\kappa_2-\kappa_1\kappa_2^3)^2+(2\kappa_1\dot\kappa_2\kappa_3+
\kappa_1\kappa_2\dot\kappa_3+3\dot\kappa_1\kappa_2\kappa_3)^2+
$$

$$\left.+
\kappa_1^2\kappa_2^2\kappa_3^2(2\kappa_2^2+\kappa_3^2+\kappa_4^2)-
2\kappa_1\kappa_2\kappa_3^2(3\ddot\kappa_1\kappa_2+\kappa_1\ddot\kappa_2+
3\dot\kappa_1\dot\kappa_2)]\right)
$$

\bigskip

\hrule

\bigskip

$$\hspace{-.5cm}
I_{10|g}={32\pi^4\over
135135}\left(221044\kappa_1^6\kappa_2^2+84050\kappa_1^4\dot\kappa_1^2+99151
\kappa_1^4\kappa_2^4-19812\kappa_1^4\kappa_2\ddot\kappa_2-
150852\kappa_1^3\ddot\kappa_1\kappa_2^2+22893\kappa_1\ddot\kappa_1\dot\kappa_1^2
-35698\kappa_1^2\dot\kappa_1\dot{\ddot\kappa}_1+\right.
$$

$$+
{13403\over 2}\dot\kappa_1^4+ {103051\over
2}\kappa_1^4\kappa_2^2\kappa_3^2-16670\kappa_1^8-23430\kappa_1^5\ddot\kappa_1+
{37557\over
2}\kappa_1^2\ddot\kappa_1^2+251264\kappa_1^3\dot\kappa_1\kappa_2\dot\kappa_2+
{63427\over
2}\kappa_1^4\dot\kappa_2^2+377962\kappa_1^2\dot\kappa_1^2\kappa_2^2+
$$

$$
+5005[(\dot{\ddot\kappa}_1-3\dot\kappa_1\kappa_2^2-
3\kappa_1\kappa_2\dot\kappa_2)^2+(3\ddot\kappa_1\kappa_2+3\dot\kappa_1\dot\kappa_2+
\kappa_1\ddot\kappa_2-\kappa_1\kappa_2^3)^2+(2\kappa_1\dot\kappa_2\kappa_3+
\kappa_1\kappa_2\dot\kappa_3+3\dot\kappa_1\kappa_2\kappa_3)^2+
$$

$$\left.+
\kappa_1^2\kappa_2^2\kappa_3^2(2\kappa_2^2+\kappa_3^2+\kappa_4^2)-
2\kappa_1\kappa_2\kappa_3^2(3\ddot\kappa_1\kappa_2+\kappa_1\ddot\kappa_2+
3\dot\kappa_1\dot\kappa_2)]\right)
$$

\bigskip

\hrule

\bigskip

$$\hspace{-.7cm}
I_{10|3}={32\pi^4\over
135135}\left(38659\kappa_1^6\kappa_2^2-677225\kappa_1^4\dot\kappa_1^2+124657
\kappa_1^4\kappa_2^4-21320\kappa_1^4\kappa_2\ddot\kappa_2-
151268\kappa_1^3\ddot\kappa_1\kappa_2^2-70616\kappa_1\ddot\kappa_1\dot\kappa_1^2
-20488\kappa_1^2\dot\kappa_1\dot{\ddot\kappa}_1-\right.
$$

$$-
28353\dot\kappa_1^4+
64298\kappa_1^4\kappa_2^2\kappa_3^2-263119\kappa_1^8+182538\kappa_1^5\ddot\kappa_1-
44382\kappa_1^2\ddot\kappa_1^2+274872\kappa_1^3\dot\kappa_1\kappa_2\dot\kappa_2+
42978\kappa_1^4\dot\kappa_2^2+458198\kappa_1^2\dot\kappa_1^2\kappa_2^2+
$$

$$
+3718[(\dot{\ddot\kappa}_1-3\dot\kappa_1\kappa_2^2-
3\kappa_1\kappa_2\dot\kappa_2)^2+(3\ddot\kappa_1\kappa_2+3\dot\kappa_1\dot\kappa_2+
\kappa_1\ddot\kappa_2-\kappa_1\kappa_2^3)^2+(2\kappa_1\dot\kappa_2\kappa_3+
\kappa_1\kappa_2\dot\kappa_3+3\dot\kappa_1\kappa_2\kappa_3)^2+
$$

$$\left.+
\kappa_1^2\kappa_2^2\kappa_3^2(2\kappa_2^2+\kappa_3^2+\kappa_4^2)-
2\kappa_1\kappa_2\kappa_3^2(3\ddot\kappa_1\kappa_2+\kappa_1\ddot\kappa_2+
3\dot\kappa_1\dot\kappa_2)]\right)
$$


\newpage

\section*{Appendix: MAPLE programs}

This Appendix contains a MAPLE program that allows one to obtain the
quantities listed in the tables with the help of a computer.

\subsection*{MAPLE program}

The MAPLE program we used in calculations looks as follows (note
that $uu[i,j]$ in the program is nothing but $U_{ij}$, $A[p,q]$ is
$A^p_qU^{2p-q+1}$, $k[i](z)$ is $\kappa_i(\tau)$ etc.; $P[d,m]$'s
are the angular integrals, formula (\ref{intgen})):

\begin{verbatim}


> #input I -> L then input P and apply L to P then input uu(k) and convert into k
>
> #  u[i] = d^iu:  uu[i,j] = (d^iu d^ju):
>
>
> M:=4: J:=3:
> A:=array(0..M,0..M): B:=array(0..M,0..M): II:=array(0..J,0..2*M+2): GG:=array(0..M,0..M):
> k:=array(0..M): K:=array(0..M): N:=array(0..M):  tu:=array(0..M): u:= array(0..M):
> uu:=array(symmetric,0..M,0..M): v:=array(0..M): VV:=array(0..M,0..M):
> for i from 0 to M do: unassign('u[i]','uu[i,i]','tu[i]'):
> for j from 0 to i-1 do:
> unassign('uu[j,i]', 'VV[j,i]'):
> od: od:
>
> #---------------------------- Parameters A[p,q] -------------------------------
> VV[0,0]:=1/V(y):
> for p from 0 to M-1 do:
> VV[p+1,0]:=diff(VV[p,0],y)/V(y):
> for q from 0 to p do:
> VV[p+1,q+1]:=(diff(VV[p,q+1],y) + VV[p,q])/V(y):
> od: od:
> V(y):=add(v[n]*y^n/n!,n=0..M):
> for p from 0 to M do: for q from 0 to M do:
> A[p,q]:=simplify(v[0]^(2*p+1-q)*eval(VV[p,q],y=0)):
> od:od:
>
> #------- list entries of TABLE II----------------------
> for p to M do
> for j from 0 to p do
> print(A[p,j]):
> od: od:
>
>
> #---entries of TABLE III: Polynomials P[d,m]-----------
>
> x:=t*t + a[0]*u+add(a[l1]*u[l1],l1=1..M): xu:= t*tu[0]+add(a[l2]*uu[0,l2],l2=0..M):
> x2:=t^2*t2 + a[0]^2 + add(a[n1]*a[n1]*uu[n1,n1],n1=1..M)+add(2*t*a[l3]*tu[l3],l3=0..M)+
> 2*add(add(a[n]*a[l4]*uu[n,l4],n=0..l4-1),l4=1..M): xL:= t*Lt+a[0]*Lu+add(a[l5]*Lu[l5],l5=1..M):
>
> x; xu; x2; xL; #----- print for illustration purposes ------
>
>
> for p to M do
> for l to p do
> P[2*p+2,2*l]:=2^(p+1)*Pi^p*2^(p-1+l)*(p-1+l)!/(2*p-1+2*l)!*add((-1)^n*(2*l)!*
> (xu)^(2*l-2*n)*(x2)^n/2^n/n!/(2*l-2*n)!*2^(p-1+2*l-n)*(p-1+2*l-n)!/2^(p-1+l)/(p-1+l)!,n=0..l):
> od: od:
>
> for p to M do
> for l from 0 to p do
> P[2*p+2,2*l+1]:=2^(p+1)*Pi^p*2^(p-1+l)*(p-1+l)!/(2*p-1+2*l)!*add((-1)^n*(2*l+1)!*
> (xu)^(2*l+1-2*n)*(x2)^n/2^n/n!/(2*l+1-2*n)!*2^(l-n)*(p+2*l-n)!/(p+l)!,n=0..l):
> od: od:
>
> #------------G[d,q,r]---------------------
>
>  #----------expect long calculation----------
> for p to M do
>   for q from 0 to p do
>   for r from 0 to p do
> JJ:=A[p,q]*A[p,r]*P[2*p+2,2*p+1-q-r]:
>  for j from 0 to M do
> LLL:=add(ii!*coeff(coeff(JJ,v[j],ii),a[j],ii),ii=0..2*M):
>  JJ:=LLL:
>  od:
> G[2*p+2,q,r]:=diff(LLL,t)/(2*p+1-q-r)!:
>  od: od:
> od:
>  #---------------expect long calculation-----------------------
>
> #---relations between uu-------
>
>    uu[0,0]:=1:
>    for j to M do
>    uuu:=add(j!*uu[n,j-n]/n!/(j-n)!,n=1..j-1):uu[0,j]:= -uuu/2:
>    uu[0,j]:= -uuu/2:
>    od:
>
> #-----entries of TABLE V: intensities of radiation of spin-J particles
>
> for p to M do
>  for q from 0 to p do
>  for r from 0 to p do
>  GG[q,r]:=simplify(G[2*p+2,q,r]): #print(simplify(G[2*p+2,q,r]),2*p+2,q,r,GG[q,r]):
>  od: od:
> II[0,2*p+2]:=GG[0,0]:
>   for j to J do
>    for q from 0 to p do
>    for r from 0 to p do
>    GG[q,r]:= add(add( uu[q1,r1]*GG[q+q1,r+r1],q1=0..p-q),r1=0..p-r):
>    unassign('q1','r1'):
>    od: od:
>   II[j,2*p+2]:=simplify(GG[0,0]):
>   od:
> od:
>
> for p to M do
> for j from 0 to J do
> print(2*p+2,j,II[j,2*p+2]):
> od: od:
>
> #--------------FRENET CALCULATION (TABLE IV)------------
>  #---------------expect long calculation-----------------------
> uu[0,0]:=1:uu[1,0]:=0:uu[1,1]:=-k[1](z)^2: N[0]:=u[0]:N[1]:=1/k[1](z)*u[1]:
> k[0](z):=1: K[0](z):=1:K[1](z):=k[1](z):
> for p from 1 to M-1 do:
> uu[p,p+1]:= diff(uu[p,p],z)/2:
> K[p+1](z):=-K[p](z)*k[p+1](z):
>  for i from 0 to p-1 do:
>  uu[i,p+1]:=-uu[i+1,p]+diff(uu[i,p],z)
>  od:
>  B[p+1,0]:=uu[0,p+1]:
>   for i from 1 to p do
>    B[p+1,i]:=-add(uu[jj,p+1]*diff(N[i],u[jj]),jj=0..p):
>   od:
>  N[p+1]:=(u[p+1]-add(B[p+1,ii]*N[ii],ii=0..p))/K[p+1](z):
>   uu[p+1,p+1]:= uu[0,p+1]^2 - K[p+1](z)^2 -add(B[p+1,ii]^2,ii=1..p):od:
>  #---------------expect long calculation-----------------------
>
>    #----list entries of TABLE IV--------------------
> for p from 0 to M do
> for j from 0 to p do uuu[p,j]:=simplify(uu[p,j]):
> print(uuu[p,j]):
> od: NN[p]:=simplify(N[p]):print(p,NN[p]): od:
> #---------------------end of FRENET-----------------------------------
>
> #-------------------intensities through FRENET (TABLE V)--------
> for p to M do
> for j from 0 to J do
> III[j,2*p+2]:=simplify(II[j,2*p+2]):
> print(2*p+2,j,III[j,2*p+2]):
> od: od:
> #--------------P_0 -- projection to u (TABLE V)--------
> for p to M do
> for j from 0 to J do
> IIp[j,2*p+2]:=subs(tu[1]=0,tu[2]=0,tu[3]=0,tu[4]=0,III[j,2*p+2]):
> print(2*p+2,j,IIp[j,2*p+2]):
> od: od:
> #--- contraction with u: t=u, tu=uu (TABLE V)
>
> for j from 0 to M do
> tu[j]:=uu[0,j]:
> od:
> for p to M do
> for j from 0 to J do
> IIII[j,2*p+2]:=simplify(II[j,2*p+2]):
> print(2*p+2,j,IIII[j,2*p+2]):
> od: od:
> for j from 0 to M do
> unassign('tu[j]'):
> od:
> #------------------------------end----------------


\end{verbatim}

\subsection*{Frenet basis}

A few comments are needed to explain how the Frenet curvatures are
handled with the program.

We deal with the Frenet basis recursively. Our main objects are the
scalars $U_{ij}$, the coefficients $D_{ij}$ of expanding Frenet
basis, ${\vec N}^{(i)}\equiv\sum_{k=0}^i D_{ik}\partial_\tau^k \vec
u$ and the coefficients $B_{ij}$ of the inverse expansion
$\partial_\tau^{j}\vec u\equiv\sum_{i=0}^{j}A_{j,i}{\vec N}^{(i)}$.
All these quantities are dealt with as functions of the Frenet
curvatures $\kappa_i$ and their derivatives.

Now, the knowledge of these quantities for all $i,j\le p$ allows one
to calculate them for all $i,j\le p+1$ in the following way:

\begin{itemize}
\item Calculate $B_{p+1,p+1}=-\kappa_{p+1} B_{p,p}$.
To prove this formula, one suffices to note that $\partial_\tau{\vec
N}^{(p)}={1\over B_{p,p}}\partial_\tau^{p+1}\vec u+\ldots\ $, which
has to be expanded into $\kappa_p{\vec N}^{(p-1)}-\kappa_{p+1}{\vec
N}^{(p+1)}\ $. However, $\partial_\tau^{p+1}\vec u\ $ contributes to
${\vec N}^{(p+1)}$ only. Therefore, $\partial_\tau{\vec
N}^{(p)}={B_{p+1,p+1}\over B_{p,p}}{\vec N}^{(p+1)}+\ldots\ $ and
$\kappa_{p+1}=-{B_{p+1,p+1}\over B_{p,p}}\ $.

\item Calculate $B_{i,p+1}$ for all $0\le i\le p-1$ using
$$
U_{i,p+1}=\partial_\tau U_{i,p}-U_{i+1,p}
$$
and $U_{p,p+1}={1\over 2}\partial_\tau U_{pp}$.
\item Calculate $B_{i,p+1}$ for all $1\le i\le p$ using
$B_{i,p+1}=-{\vec N}^{(i)}\partial_\tau^{p+1}\vec u=-\sum_k
D_{ik}U_{k,p+1}$, the latter expression can be simbolically written
as $\left\{-\sum_k{\partial {\vec N}^{(i)}\over\partial
(\partial_\tau^k \vec u)}U_{k,p+1}\right\}$. Note that
$B_{p+1,0}=U_{0,p+1}$.
\item Calculate ${\vec N}^{(p+1)}={\partial_\tau^{p+1}\vec u-\sum_{j=0}^p
B_{p+1,j}{\vec N}^{(j)}\over B_{p+1,p+1}}$.
\item Calculate
$U_{p+1,p+1}=\left(B_{p+1,0}\right)^2-\sum_{i=1}^{p+1}\left(B_{p+1,i}\right)^2$.
\end{itemize}

\end{document}